\documentclass[prb, twocolumn, amsmath, amssymb, superscriptaddress]{revtex4-2}
\usepackage{graphicx}
\usepackage{color}

\usepackage{amsmath,amssymb,bm}

\usepackage{physics}
\usepackage{siunitx}
\usepackage{times}
\usepackage{hyperref}
\AtBeginDocument{\usepackage{booktabs}}

\date{\today} 

\begin{document}
\title{Density matrix renormalization group study of the interacting Kitaev chain with quasi-periodic disorder}

\author{K.~S.~C.~Decker}
\affiliation{Technische Universit\"at Braunschweig, Institut f\"ur Mathematische Physik, Mendelssohnstraße 3, 38106 Braunschweig, Germany}

\author{C.~Karrasch}
\affiliation{Technische Universit\"at Braunschweig, Institut f\"ur Mathematische Physik, Mendelssohnstraße 3, 38106 Braunschweig, Germany}

\begin{abstract}
We document the ground state phase diagram of the one-dimensional Kitaev chain with quasi-periodic disorder in the presence of two-body interactions. Our data was obtained for systems of $L=1000$ sites using large-scale density matrix renormalization group numerics and is benchmarked against known results for the clean system. We demonstrate that moderate quasi-periodic disorder stabilizes the topological phase both for repulsive and attractive interactions. For larger disorder strengths, the system features re-entrance behavior and multiple phase transitions.
\end{abstract}

\maketitle

The Kitaev chain is a drosophila model to study topological phases in one dimension \cite{Kitaev2001}. It is known to host Majorana edge modes, which are of interest in the context of quantum computation \cite{Deutsch1985,Ladd2010,Nielsen2010,Nayak2008}. The quest for experimental observations of the existence of the Majorana zero-modes has been an elusive one, see, e.g., \cite{KT,doi:10.1126/science.1222360,Nadj2014,kouwmajo,hbmajo}.

While it is guaranteed that the topological phase of the Kitaev chain is protected against small symmetry-preserving perturbations, the phase diagram away from this limit needs to be established explicitly. In the presence of two-body interactions, novel phases (such as Mott-insulating charge-density waves) emerge, and consensus about the phase diagram has been largely reached in the literature \cite{Hassler_2012,PhysRevB.84.085114,PhysRevB.88.161103,Katsura2015,Mahyaeh2020,PhysRevLett.118.267701,PhysRevB.92.104514}. The effects of disorder on the non-interacting system was studied extensively \cite{DeGottardi2013,DeGottardi2013b,Altland2014,PhysRevB.107.165137}. Questions about topological phases and Majorana edge modes in the presence of interactions and/or disorder were also addressed in a variety of other models \cite{Lobos2012,Crepin2014,PhysRevLett.107.036801,PhysRevB.84.014503,PhysRevB.84.094503,PhysRevB.84.214528,Manolescu2014,PhysRevLett.115.166401,PhysRevB.92.081401,PhysRevB.92.085139,PhysRevB.92.155434,PhysRevB.92.235123,Brouwer2011,Pientka2013,Lobos2012,Crepin2014,PhysRevB.103.195119,PhysRevB.106.224505,PhysRevB.107.014202,PhysRevLett.131.186303,PhysRevResearch.4.L032016,10.21468/SciPostPhys.14.6.152,PhysRevB.100.134207,PhysRevB.107.155134,PhysRevB.106.184505,PhysRevB.107.125110}.

The combined effect of interactions and quenched or Fibonacci-type disorder on the ground state of the Kitaev chain was studied in a few works \cite{Gergs2016,PhysRevB.96.241113,Crepin2014,Lobos2012,PhysRevB.97.085425,universe5010033,PhysRevB.105.245144}, e.g., using density-matrix renormalization group (DMRG) numerics \cite{Gergs2016,PhysRevB.105.245144}. It was shown that the topological phase is enlarged by moderate quenched disorder or moderate repulsive interactions. Quasi-periodic potentials (see Eq.~(\ref{eq:dis})) have generally received wide-spread attention in the past, partly due to the possibility that they can be implemented in cold atom setups (for an example from the world of many-body localization see, e.g., Ref.~\onlinecite{doi:10.1126/science.aaa7432}). Within the realm of the Kitaev chain, quasi-periodic disorder has been investigated in the non-interacting limit \cite{DeGottardi2013,PhysRevB.86.205135,PhysRevLett.110.176403} or for spinful generalizations featuring Hubbard-$U$ interactions \cite{Tezuka2012}.

The key goal of our paper is to extend these studies and to establish the phase diagram of the Kitaev chain with quasi-periodic disorder and nearest-neighbor interactions for systems of $L=1000$ sites. Our data was obtained using large-scale density-matrix renormalization group calculations with a total run-time (including for data not shown) of approximately 30,000 core-hours. Figure \ref{fig:phase2} is the main result. We revisit the issue of how the different phases can be detected numerically \cite{Gergs2016} and carefully investigate that our method yields converged results. The clean limit, where consensus about the phase diagram has largely been reached in the literature, serves as a testing ground.

\section{Model}

The one-dimensional, interacting, disordered Kitaev chain is governed by 
\begin{align}
    \mathcal{H} = &\sum_{j=1}^{L-1} \big[ -t c_{j}^{\dagger} c_{j+1} + \delta c_{j} c_{j+1} + \text{h.c.} \nonumber\\
    & + U (2 n_{j} - 1) (2 n_{j+1} - 1)\big]
    -  \sum_{j=1}^L\mu_j  n_{j},\label{eq:H1}
\end{align}
where $c_{j}^{\dagger}$ and $c_{j}$ are the creation and annihilation operators for spinless fermions at site $j$, and $n_{j}=c_{j}^{\dagger} c_{j}$ is the number operator. The hopping amplitude and the superconducting gap are denoted by $t$ and $\delta\in\mathbb{R}$, respectively. We introduce a nearest-neighbor interaction $U$ as well as a quasi-periodic chemical potential of strength $\Delta$
\begin{equation}\label{eq:dis}
\mu_j = \mu + \Delta \cos(2\pi\beta j+\phi).
\end{equation}
The results are expected not to depend on the (irrational) value of $\beta$, and we choose the golden ratio $\beta=(1+\sqrt{5})/2$ that is usually employed in the literature (see the above references). We mainly choose a phase $\phi=0$ but will exemplify that averaging over $\phi$ yields identical results. The system size is $L$, and we employ open boundary conditions (OBC). The signs of $t$, $\delta$, and (for $\Delta=0$) also $\mu$ are irrelevant, which follows from $c_j\to i(-1)^j c_j$, $c_j\to ic_j$, and $c_j\to(-1)^jc_j^\dagger$, respectively. In the following, we will always choose $t>0$ and $\delta>0$.

For the purpose of treating the system with the density matrix renormalization group, it is advantageous to rewrite the Hamiltonian in terms of a spin chain by virtue of a Jordan-Wigner transformation
\begin{equation}\begin{split}
\sigma^{x}_{j} & = \phantom{-i}\prod_{l<j}(1-2 n_{l})(c_{j}^{\dagger} + c_{j}), ~~ \sigma^{z}_{j} = 2n_{j}-1,\\
\sigma^{y}_{j} & = -i\prod_{l<j}(1-2 n_{l})(c_{j}^{\dagger} - c_{j}),
\end{split}\end{equation}
where $\sigma^{x,y,z}_j$ represent the usual Pauli matrices \cite{Jordan1928,Fendley2012}. This yields
\begin{align}
    \mathcal{H} = - \sum_{j=1}^{L-1} \bigg[ \frac{(t + \delta)}{2}  & \sigma^{x}_{j} \sigma^{x}_{j+1}  + \frac{(t - \delta)}{2} \sigma^{y}_{j} \sigma^{y}_{j+1} \nonumber \\ - U & \sigma^{z}_{j} \sigma^{z}_{j+1} \bigg]
    -\frac{1}{2}\sum_{j=1}^L \mu_{j} \sigma^{z}_{j}.\label{eq:H3}
\end{align}
In the spin language, the invariance under the sign flip of $t$, $\delta$, and $\mu$ follows from $\sigma_j^{x,y}\to \pm (-1)^j\sigma_j^{y,x}$, $\sigma_j^{x,y}\to \pm \sigma_j^{y,x}$, and $\sigma_j^{x,z}\to-\sigma_j^{x,z}$, respectively.

\section{Simple limits}

It is instructive to recapitulate the phase diagram of the Kitaev chain in several exactly-solvable limits. This will aid the definition of proper probes to identify the various phases. We include an extensive discussion both for didactical purposes and in order to make this paper self-contained; a knowledgeable reader can skip directly to Sec.~\ref{sec:method}.

\subsection{Non-interacting, clean case}

In the absence of interactions and disorder ($U=\Delta=0$), the Hamiltonian (\ref{eq:H1}) with periodic boundary conditions (PBC) and $L$ even can be diagonalized straightforwardly by going to momentum space $c_k=\frac{1}{\sqrt{L}}\sum_j e^{ikj}c_j$, $k=\frac{2\pi}{L}n$, $n=-\frac{L}{2}+1,...,\frac{L}{2}$. If one subsequently employs a Bogoliubov transformation $\eta_k=v_kc_k-iv_kc_{-k}^\dagger$ for $k\notin\{0,\pi\}$, one obtains (up to a constant) \cite{Fendley2012,Calabrese_2012}
\begin{equation}\label{eq:Hs}
\mathcal{H}_\text{Kit} = \sum_{\genfrac{}{}{0pt}{2}{k\neq0}{k \neq\pi}} E_k\Big(\eta_k^\dagger\eta_k -\frac{1}{2}\Big) + \epsilon_0\Big(c_0^\dagger c_0-\frac{1}{2}\Big)+\epsilon_\pi\Big(c_\pi^\dagger c_\pi-\frac{1}{2}\Big),
\end{equation}
with $\epsilon_k=-2t\cos k -\mu$, $\delta_k=2\delta\sin k$, and $E_k=\sqrt{\epsilon_k^2 + \delta_k^2}$. For $\mu<0$, the ground state is determined by $\eta_k|\text{vac}\rangle=c_\pi|\text{vac}\rangle=0$, and the zero-momentum mode is unoccupied for $\mu<-2t$ and occupied otherwise ($\mu>0$ follows analogously). The system becomes gapless for $|\mu|=2t$ but is gapped away from these points.

Note that if PBC are imposed directly on the fermionic Hamiltonian (\ref{eq:H1}), the ground state is always non-degenerate even in the thermodynamic limit. In the case of open boundary conditions, the ground state becomes two-fold degenerate for $|\mu|<2t$ and $L\to\infty$. It is instructive to recall that Eq.~(\ref{eq:H1}) maps to Eq.~(\ref{eq:H3}) only for OBC; due to the non-local nature of the Jordan-Wigner transformation, PBC in the spin model map to PBC or anti-PBC in the fermionic language depending on the parity, and the ground state is two-fold degenerate for $|\mu|<2t$ and $L\to\infty$.

Since a gap-closing occurs only for $|\mu|=2t$, it is immediately clear that the phases at different $\delta>0$ are adiabatically connected (at least for $U=\Delta=0$), and it is thus reasonable to focus on a single value of $\delta$. If one sets $\delta=t$, the discussion of the non-interacting, clean limit becomes particularly simple since one recovers the well-known \textit{transverse-field Ising chain} \cite{SachdevBook}:
\begin{equation}\label{eq:Htfi}
\mathcal{H}_\text{TFI} = -\sum_{j} \Big(t \sigma^{x}_{j} \sigma^{x}_{j+1} + \frac{\mu}{2} \sigma^{z}_{j}\Big).
\end{equation}
For $\abs{\mu} < 2t$, the system is in a ferromagnetic (FM) phase with an order parameter $\langle \sigma^{x}_{j} \rangle \neq 0$.
In the fermionic language, this regime corresponds to a topological phase with a non-local order parameter \cite{Fendley2012}. For $\abs{\mu} > 2t$ the system is in a paramagnetic (PM) phase with $\langle \sigma^{x}_{j} \rangle = 0$, which corresponds to a fermionic band insulator.

\subsection{XY chain}

In the absence of a chemical potential ($\mu=\Delta=0$) and for $\delta=t$, one recovers the XY model, which can be mapped to free fermions $\tilde c_j$ simply by switching labels \cite{LSM1961}:
\begin{equation}\label{eq:Hxy}\begin{split}
\mathcal{H}_\text{XY} & = \sum_{j} \big(-t \sigma^{x}_{j} \sigma^{x}_{j+1} + U \sigma^{z}_{j}\sigma^{z}_{j+1}\big)\\
& = \sum_j\Big[ (U-t) \tilde c_j^\dagger \tilde c_{j+1} + (U+t) \tilde c_j \tilde c_{j+1} + \text{h.c.} \Big].
\end{split}\end{equation}
The sign of both $t$ and $U$ is irrelevant. The system is in a gapped phase for $|U|<t$ which is adiabatically connected to the FM phase of the Ising chain with $\langle\sigma^x_j\rangle\neq0$. The gap closes at the isotropic points $|U|=t$. For $|U|>t$, the system is again gapped. It features ferromagnetic order with $\langle\sigma^z_j\rangle\neq0$ for $U<-t$, which in the realm of the original model is adiabatically connected to the trivial insulator. At $U>t$, it exhibits anti-ferromagnetic order $\langle\sigma^z_j\rangle\sim (-1)^jN_0$ for $U>t$ (`Mott insulator'), which can be seen easily by virtue of $\sigma_j^{y,z}\to (-1)^j\sigma_j^{y,z}$.

\begin{figure}[t!]
\includegraphics[width=0.9\linewidth,clip]{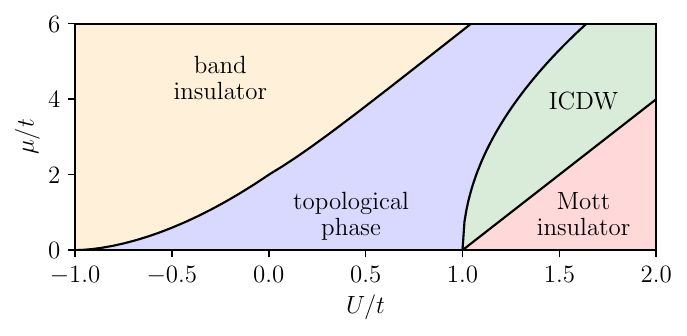}
\caption{Sketch of phase diagram of the interacting Kitaev chain with $\delta=t$ in the absence of disorder ($\Delta=0$), following Refs.~\onlinecite{Hassler_2012,Katsura2015}.}
\label{fig:sketch}
\end{figure}

\subsection{Duality}

If one sets $\delta=t$ and $\Delta=0$ and performs a non-local duality transformation $\tau_{j}^{z} = \sigma_{j}^{x} \sigma_{j+1}^{x}$ and $\tau_{j}^{x} = \prod_{l < j} \sigma_{l}^{z}$, the Kitaev chain can be mapped onto the axial next-nearest-neighbor Ising (ANNNI) model \cite{Selke1988,Beccaria2007}:
\begin{equation}\begin{split}\label{eq:Hannni}
    \mathcal{H}_{\text{ANNNI}} &=
\sum_{j} \big(-t \sigma^{x}_{j} \sigma^{x}_{j+1} + U \sigma^{z}_{j}\sigma^{z}_{j+1} -\frac{\mu}{2}\sigma_j^z \big) \\ & =
     \sum_{j}\Big(  - \frac{\mu}{2} \tau_{j}^{x} \tau_{j+1}^{x}
    + U \tau_{j}^{x} \tau_{j+2}^{x} -t\tau_{j}^{z}\Big).
\end{split}\end{equation}
For $U=0$, one recovers the self-duality of the transverse-field Ising chain. Note that the PM and FM phases are interchanged, which illustrates that due to the non-locality of the duality transformation, properties such as ground-state degeneracies are not in a one-to-one correspondence if the influence of boundary effects is disregarded. This does not hold true for the spectral gap, and phases that are adiabatically connected in the dual model are also adiabatically connected in the original one.

Thus, one can exploit the fact that the phase diagram of the ANNNI model has been studied extensively in the literature \cite{Hassler_2012,Katsura2015,Mahyaeh2020}, e.g., using analytical arguments or numerical approaches \cite{Peschel1981,Selke1988,Allen2001,Beccaria2006,Beccaria2007,Sela2011}. In addition to the FM and PM phases, it features an antiphase of the form $\uparrow_x\uparrow_x\downarrow_x\downarrow_x$ that appears for large $U$ and that corresponds to the Mott insulator in the original language. Moreover, a novel floating phase emerges in between the PM phase and the antiphase; for the fermions, this corresponds to an incommensurate charge density wave (ICDW). Note that while the FM/PM phase boundary as well as the existence of the antiphase at large $U$ are well established, the specifics of the intermediate regime are still being debated. Since the main focus of this work is on the effects of disorder, we refrain from delving into these issues.

If one recasts the phase diagram of the ANNNI model in terms of the `non-standard' axis $\mu/t$ and $U/t$, one obtains the sketch shown in Fig.~\ref{fig:sketch}, see Refs.~\onlinecite{Hassler_2012,Katsura2015}. One can readily identify the critical points i) $|\mu|=2t$ at $U=0$, which is associated with the FM/PM transition in the ANNNI model as well as ii) $|U|=t$ at $\mu=0$, where the ANNNI model reduces to two independent Ising chains and which is associated with the FM/PM transition at $U=-t$ and with a multi-critical point at $U=t$ where all four phases meet \cite{PhysRevLett.118.267701}. Note that the conventional multi-critical point $U=\mu/4$ at $t=0$ is not realized in our case (since the fermionic hopping is always finite).

\begin{figure}[t]
\includegraphics[width=\linewidth,clip]{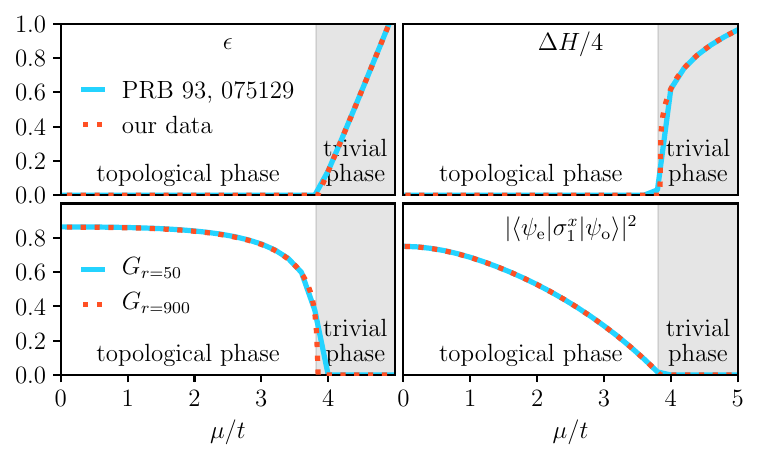}
\caption{Comparison of our DMRG data for $\Delta=0$ (no disorder), $\delta=t$, $U/t=0.5$, $L=1000$ with the results of Ref.~\onlinecite{Gergs2016} that were obtained for $L=200$ and using a different approach to determine degenerate ground states. The energy separation $\epsilon$, the entanglement gap $\Delta H$, the correlation function $G_r$, and the Majorana overlap $\langle\psi_\text{e}|\sigma_1^x|\psi_\text{o}\rangle$ can be used to identify the topological and (trivial) band insulating phases, respectively.}
\label{fig:test}
\end{figure}

\section{Method}
\label{sec:method}

We determine the ground state of the system variationally using the density matrix renormalization group in a matrix-product state formulation \cite{White1992,Schollwoeck2005,Schollwoeck2011}. We work with OBC and a fixed system size of $L=1000$. All data in this work was computed using a two-site DMRG algorithm; we have checked that a one-site algorithm yields identical results. The initial state is determined using the infinite-system DMRG for the homogeneous system.

The bond dimension $\chi$ is the key numerical control parameter; in order to test convergence, we compute the re-scaled energy variance
\begin{equation}
    \frac{\Delta E}{E^2} = \frac{\langle \psi | \mathcal{H}^{2} | \psi \rangle - \langle \psi | \mathcal{H} | \psi \rangle^{2}}{\langle \psi | \mathcal{H} | \psi \rangle^{2}}~.
\end{equation}
We successively increase the bond dimension $\chi=12\to24\to48$ after 20 DMRG sweeps each. It turns out that $\chi=48$ is sufficient in order for $\Delta E/E^2$ to drop to machine precision except in the incommensurate CDW phase that appears in the absence of disorder (which is not at the focus of this work). The Hamiltonian can be expressed as a matrix-product operator with a bond dimension of $\chi_{\mathcal H}=5$ in the most general case, $\chi_{\mathcal H}=4$ at $\delta=t$ or $U=0$, and $\chi_{\mathcal H}=3$ if both $\delta=t$ and $U=0$.

After the ground state $|\psi_0\rangle$ of $\mathcal{H}$ has been obtained, we can calculate low-lying states by computing the ground state $|\psi_1\rangle$ of
\begin{equation}
\mathcal{H}'=\mathcal{H}+h|\psi_0\rangle\langle\psi_0|,
\end{equation}
where we typically work with an energy penalty $h=10t$. This procedure can be implemented straightforwardly using matrix product states.

In order to detect the different phases such as the trivial insulator, the topological phase, and the charge density wave phase, we employ various measures \cite{Gergs2016}. The difference between the energies $E_{0,1}$ of the two lowest-lying states
\begin{equation}
\epsilon = 2 \frac{E_{1} - E_{0}}{E_{1} + E_{0}}
\end{equation}
can be used to identify degenerate ground states. Note that while $\epsilon>0$ necessarily implies the existence of a gap, the opposite is not true if the ground state is degenerate. We call the ground degenerate if $\epsilon<10^{-9}$. The DMRG typically converges into lowly-entangled states, and we introduce $|\psi_\text{e,o}\rangle=(|\psi_0\rangle\pm|\psi_1\rangle)/\sqrt{2}$.

\begin{figure*}[t]
\includegraphics[width=\linewidth,clip]{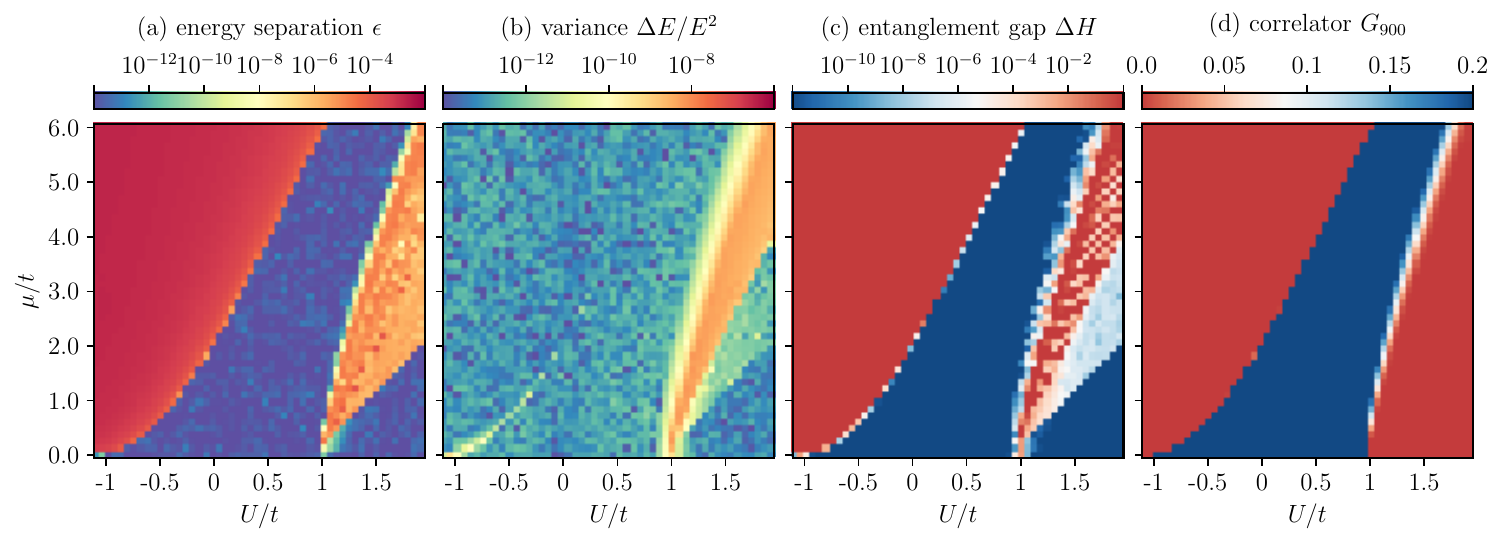}
\caption{Phase diagram of the Kitaev chain in the absence of disorder ($\Delta=0$) and $\delta=t$ as a function of the chemical potential $\mu$ and the interaction $U$. This regime has been widely investigated in the literature (see the main text) and serves as a testing ground for our numerics. The trivial band insulator, the topological phase, and the charge density wave manifest in the top left, center, and bottom right corners, respectively (see the main text for details). The results were calculated for a system of $L=1000$ sites using a 2-site DMRG algorithm with a final bond dimension of $\chi=48$. This is sufficient to obtain results up to machine precision (variance $\Delta E/E^2\sim10^{-12}$) except in regions where prior works predict an incommensurate CDW, which is not at the focus of this work.}
\label{fig:phase1a}
\end{figure*}

Another probe is the degeneracy of the bipartite entanglement spectrum \cite{PhysRevB.84.014503,PhysRevB.83.075102,RevModPhys.81.865}, which we probe via the gap $\Delta H$ in the spectrum of $-\text{log\,Tr}_{L/2}|\psi_0\rangle\langle\psi_0|$, where $|\psi_0\rangle$ is the ground state. In the case of degenerate ground states, we compute $\Delta H$ in an equal superposition $|\psi_\text{e}\rangle$ of these states. Moreover, the correlator
\begin{equation}
    G_{r} = \frac{1}{L-r} \sum_{j=1}^{L-r} \abs{\langle \sigma_{j}^{x} \sigma_{j+r}^{x} \rangle}
\end{equation}
is known to be finite (zero) in the topological (band insulating) phase \cite{Gergs2016,Fendley2012}, which is plausible since $\langle\sigma_j^x\rangle$ is just the order parameter of the transverse-field Ising chain in the spin language. Since our system is disordered, we set $r=0.9L$, which entails an average over $L-r=0.1L$ different values of the correlator. In the case of degenerate ground states, we compute $G_r$ in the first state that we obtain. The topological phase can also be detected from the overlap $\langle\psi_\text{e}|\sigma_1^x|\psi_\text{o}\rangle$, which directly probes the existence of Majorana edge modes \cite{PhysRevB.84.014503,Gergs2016}. Note that since prior works concluded that the Majonara overlap is not particularly suitable \cite{Gergs2016}, it is not at the focus of this paper, and for simplicity we just compute the quantity $\langle\psi_\text{e}|\sigma_1^x|\psi_\text{o}\rangle$ as a check.

\begin{figure}[b]
\includegraphics[width=\linewidth,clip]{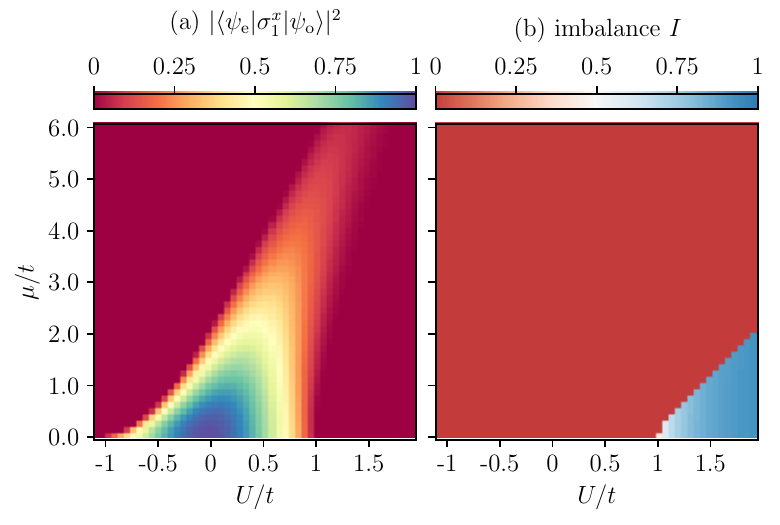}
\caption{The same as in Fig.~\ref{fig:phase1a} but for the Majorana overlap and the charge imbalance.}
\label{fig:phase1b}
\end{figure}

\begin{figure*}[t]
\includegraphics[width=\linewidth,clip]{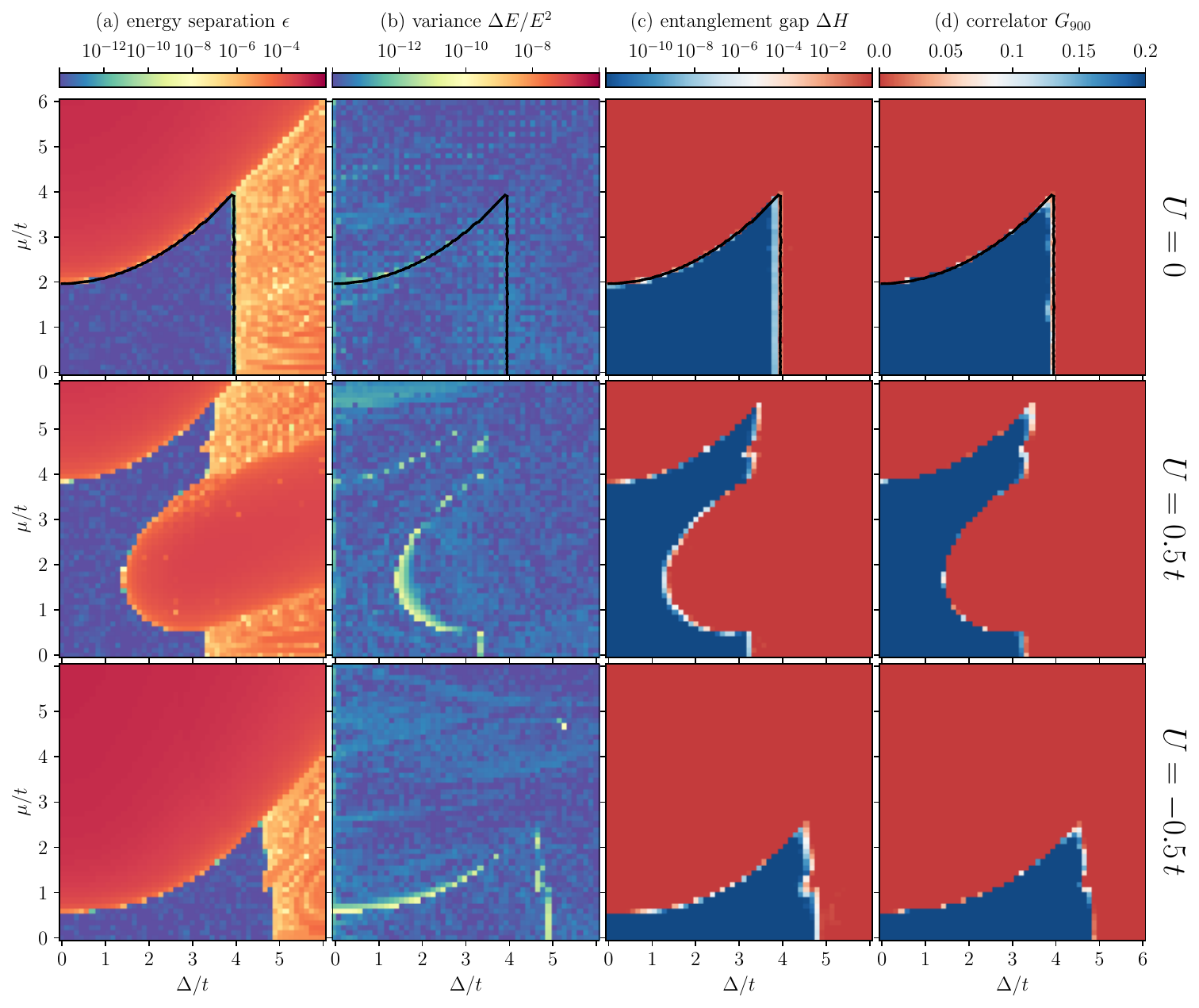}
\caption{Phase diagram of the Kitaev chain with $L=1000$ as a function of the chemical potential $\mu$ and the strength of the quasi-periodic disorder $\Delta$ for $\delta=t$ and various interactions $U/t\in\{0,\pm0.5\}$. Red and blue regions in (a), (c), and (d) represent the trivial band insulating and topological phase, respectively. The solid black lines shows the boundary between the topological and band insulating phases at $U=0$ computed by diagonalizing the non-interacting single-particle problem. }
\label{fig:phase2}
\end{figure*}

\section{Results}
\label{sec:results}

We first compare our results in the absence of disorder ($\Delta=0$) with those of Ref.~\onlinecite{Gergs2016}. This is instructive not only as a benchmark but also because degenerate ground states were determined differently in this prior work, namely by introducing a small field in $\sigma^x$-direction instead of replacing $\mathcal{H}\to\mathcal{H}'$ \cite{dirkcomment}. The result is shown in Fig.~\ref{fig:test} where we compare the energy separation $\epsilon$, the entanglement gap $\Delta H$, the correlation function $G_r$, and the Majorana overlap $\langle\psi_\text{e}|\sigma_1^x|\psi_\text{o}\rangle$. All four quantities detect the transition from the topological phase into the trivial band insulating phase as the chemical potential is increased. Note that in the absence of interactions, this transition occurs at $\mu/t=2$, and the size of the topological phase is hence enlarged by a repulsive $U/t=0.5$. We quantitatively reproduce the data of Ref.~\onlinecite{Gergs2016}; minor discrepancies (see, e.g., $G_r$) are readily explained by the different system size.

As a next step, we compute the phase in the absence of disorder as a function of the chemical potential $\mu$ and the interaction $U$. Since consensus has been largely reached in the literature \cite{Hassler_2012,PhysRevB.84.085114,PhysRevB.88.161103,Katsura2015,Mahyaeh2020,PhysRevLett.118.267701,PhysRevB.92.104514}, this step will aid the interpretation of the results for finite $\Delta$. A sketch of the phase diagram, which stems from mapping the problem to the ANNNI model, is shown in Fig.~\ref{fig:sketch} following Refs.~\onlinecite{Hassler_2012,Katsura2015}.

Our DMRG data for a system of $L=1000$ sites is shown in Figs.~\ref{fig:phase1a} and \ref{fig:phase1b}. The trivial band insulator is characterized by an energy gap above a non-degenerate ground state ($\epsilon>0$), a finite entanglement gap $\Delta H>0$, and a vanishing correlator $G_r=0$. This phase manifests in the top left corner of Fig.~\ref{fig:phase1a} and is adiabatically connected to the line $U=0$, $\mu/t>2$. The topological phase features a degenerate ground state, a vanishing entanglement gap $\Delta H=0$, and a finite correlator $G_r>0$; it manifests in the center of Fig.~\ref{fig:phase1a} and is adiabatically connected to the line $U=0$, $\mu<2t$. A sharp phase boundary between the topological and trivial band insulating phases can be extracted from these quantities. The same does not hold true for the Majorana overlap $\langle\psi_\text{e}|\sigma_1^x|\psi_\text{o}\rangle$ (see Fig.~\ref{fig:phase1b}(a)), which we therefore do not employ further. Similar conclusions were already reached in Ref.~\onlinecite{Gergs2016}.

At large $U/t$ and small $\mu/t$, we observe a CDW phase governed by a degenerate ground state, a vanishing entanglement gap $\Delta H=0$, and a vanishing correlator $G_r=0$ (bottom right corner of Fig.~\ref{fig:phase1a}). This phase can be identified unambiguously from the charge imbalance
\begin{equation}
I = \abs{\frac{N_\text{a}-N_\text{b}}{N_\text{a}+N_\text{b}}},~~N_\text{a,b}=\sum_{j\text{ even, odd}}\frac{\langle\sigma^z_j\rangle+1}{2},
\end{equation}
which is shown in Fig.~\ref{fig:phase1b}(b). In the case of degenerate ground states, we compute $I$ in the first state that we obtain.

\begin{figure*}[t]
\includegraphics[width=\linewidth,clip]{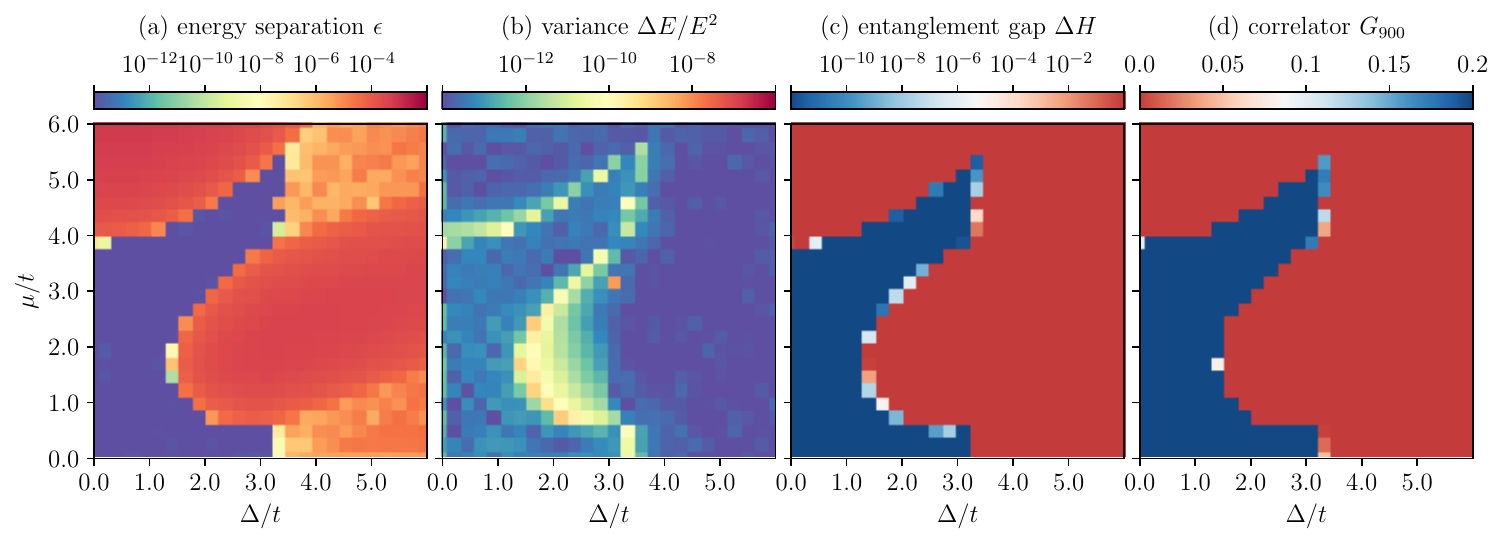}
\caption{The same phase diagram (lower resolution) as in Fig.~\ref{fig:phase2} for $U/t=0.5$ but with a disorder averaging over 8 phases $\phi = 2\pi k/n$ with $k=0\ldots n-1$ and $n=8$.}
\label{fig:phaseav}
\end{figure*}

In Fig.~\ref{fig:phase1a}(b), we show the variance $\Delta E/E^2$ of the ground state (if the ground state is degenerate, we show the larger variance). The bond dimension of $\chi=48$ is sufficient to obtain data up to machine precision ($\Delta E/E^2\sim10^{-12}$) except in a region where one expects an incommensurate charge density wave. This regime is not at the focus of this work.

Our observations in the absence of disorder are consistent with the sketch in Fig.~\ref{fig:sketch} and with the results available in the literature \cite{Hassler_2012,PhysRevB.84.085114,PhysRevB.88.161103,Katsura2015,Mahyaeh2020,PhysRevLett.118.267701,PhysRevB.92.104514}.

We now turn to the case with finite quasi-periodic disorder. The phase diagram is shown in Fig.~\ref{fig:phase2} as a function of the chemical potential $\mu$ and the disorder strength $\Delta$ for three values of the interaction $U/t\in\{0,\pm0.5\}$. In the absence of interactions, one does not need to resort to DMRG numerics but can simply diagonalize a single-particle Hamiltonian numerically. The band insulating and topological phases can be detected, e.g., by the absence or presence of zero-energy modes. The resulting phase boundary is shown as a solid black line in Fig.~\ref{fig:phase2} and is in agreement with our DMRG data; note that in this case, red and blue regions are unambiguously associated with the band insulating and topological phase, respectively. We observe that finite disorder first stabilizes the topological phase until it eventually breaks down. Similar observations were made in the case of quenched disorder \cite{Gergs2016}.

The topological phase is stabilized by repulsive interactions $U/t=0.5$ for small disorder strength and vice-versa, see Fig.~\ref{fig:phase2}. This is again consistent with the case of quenched disorder \cite{Gergs2016}. At intermediate values of $\Delta$, one observes a non-monotonous dependency w.r.t.~the chemical potential and hence re-entrance behavior and multiple phase transitions. For even larger $U/t=1.25$ (data not shown), the bottom-left corner of the phase diagram is governed by CDW behavior and disorder again stabilizes the topological phase for large values of $\mu$; we still observe non-monotonicities and expect that they manifest for all values of $U$ until eventually CDW behavior sets in.

Attractive interactions $U/t=-0.5$ destabilize the topological phase, which is again stabilized by disorder. At $U/t=-1$ (data not shown), the topological phase can be induced by disorder for small chemical potentials.

All data shown up to this point was computed for a fixed value $\phi=0$ of the phase. In Fig.~\ref{fig:phaseav}, we show results similar to those of Fig.~\ref{fig:phase2} at $U/t=0.5$ but now averaged over various $\phi$. The phase diagram does not change, which is not surprising due to the self-averaging nature of this type of disorder.

\section{Summary}

We have documented large-scale DMRG data (30,000 core hours) for the ground-state phase diagram of the Kitaev chain in the presence of nearest-neighbor interactions and quasi-periodic disorder. Moderate disorder or repulsive interactions generally stabilize the topological phase. In certain regimes, one can observe multiple transitions between the trivial band insulating and the topological phase as the chemical potential is varied. As a next step, one can envision targeting excited states using tensor networks, e.g., via the DMRG-X \cite{Khemani2016} algorithm, which would be of relevance for the realm of many-body localization.

\subsection*{Data Availability Statement}

All data are available from the authors upon reasonable request.

\subsection*{Author Contributions}

K.D. carried out the DMRG calculations. C.K. interpreted the results. Both authors worked on the manuscript.

\bibliography{references}

\begin{thebibliography}{72}%
\makeatletter
\providecommand \@ifxundefined [1]{%
 \@ifx{#1\undefined}
}%
\providecommand \@ifnum [1]{%
 \ifnum #1\expandafter \@firstoftwo
 \else \expandafter \@secondoftwo
 \fi
}%
\providecommand \@ifx [1]{%
 \ifx #1\expandafter \@firstoftwo
 \else \expandafter \@secondoftwo
 \fi
}%
\providecommand \natexlab [1]{#1}%
\providecommand \enquote  [1]{``#1''}%
\providecommand \bibnamefont  [1]{#1}%
\providecommand \bibfnamefont [1]{#1}%
\providecommand \citenamefont [1]{#1}%
\providecommand \href@noop [0]{\@secondoftwo}%
\providecommand \href [0]{\begingroup \@sanitize@url \@href}%
\providecommand \@href[1]{\@@startlink{#1}\@@href}%
\providecommand \@@href[1]{\endgroup#1\@@endlink}%
\providecommand \@sanitize@url [0]{\catcode `\\12\catcode `\$12\catcode
  `\&12\catcode `\#12\catcode `\^12\catcode `\_12\catcode `\%12\relax}%
\providecommand \@@startlink[1]{}%
\providecommand \@@endlink[0]{}%
\providecommand \url  [0]{\begingroup\@sanitize@url \@url }%
\providecommand \@url [1]{\endgroup\@href {#1}{\urlprefix }}%
\providecommand \urlprefix  [0]{URL }%
\providecommand \Eprint [0]{\href }%
\providecommand \doibase [0]{https://doi.org/}%
\providecommand \selectlanguage [0]{\@gobble}%
\providecommand \bibinfo  [0]{\@secondoftwo}%
\providecommand \bibfield  [0]{\@secondoftwo}%
\providecommand \translation [1]{[#1]}%
\providecommand \BibitemOpen [0]{}%
\providecommand \bibitemStop [0]{}%
\providecommand \bibitemNoStop [0]{.\EOS\space}%
\providecommand \EOS [0]{\spacefactor3000\relax}%
\providecommand \BibitemShut  [1]{\csname bibitem#1\endcsname}%
\let\auto@bib@innerbib\@empty
\bibitem [{\citenamefont {Kitaev}(2001)}]{Kitaev2001}%
  \BibitemOpen
  \bibfield  {author} {\bibinfo {author} {\bibfnamefont {A.~Y.}\ \bibnamefont
  {Kitaev}},\ }\bibfield  {title} {\bibinfo {title} {Unpaired majorana fermions
  in quantum wires},\ }\href {https://doi.org/10.1070/1063-7869/44/10s/s29}
  {\bibfield  {journal} {\bibinfo  {journal} {Phys.-Usp.}\ }\textbf {\bibinfo
  {volume} {44}},\ \bibinfo {pages} {131} (\bibinfo {year} {2001})}\BibitemShut
  {NoStop}%
\bibitem [{\citenamefont {Deutsch}(1985)}]{Deutsch1985}%
  \BibitemOpen
  \bibfield  {author} {\bibinfo {author} {\bibfnamefont {D.}~\bibnamefont
  {Deutsch}},\ }\bibfield  {title} {\bibinfo {title} {Quantum theory, the
  church–turing principle and the universal quantum computer},\ }\href
  {https://doi.org/10.1098/rspa.1985.0070} {\bibfield  {journal} {\bibinfo
  {journal} {Proc. R. Soc. Lond. A}\ }\textbf {\bibinfo {volume} {400}},\
  \bibinfo {pages} {97} (\bibinfo {year} {1985})}\BibitemShut {NoStop}%
\bibitem [{\citenamefont {Ladd}\ \emph {et~al.}(2010)\citenamefont {Ladd},
  \citenamefont {Jelezko}, \citenamefont {Laflamme}, \citenamefont {Nakamura},
  \citenamefont {Monroe},\ and\ \citenamefont {O'Brien}}]{Ladd2010}%
  \BibitemOpen
  \bibfield  {author} {\bibinfo {author} {\bibfnamefont {T.~D.}\ \bibnamefont
  {Ladd}}, \bibinfo {author} {\bibfnamefont {F.}~\bibnamefont {Jelezko}},
  \bibinfo {author} {\bibfnamefont {R.}~\bibnamefont {Laflamme}}, \bibinfo
  {author} {\bibfnamefont {Y.}~\bibnamefont {Nakamura}}, \bibinfo {author}
  {\bibfnamefont {C.}~\bibnamefont {Monroe}},\ and\ \bibinfo {author}
  {\bibfnamefont {J.~L.}\ \bibnamefont {O'Brien}},\ }\bibfield  {title}
  {\bibinfo {title} {Quantum computers},\ }\href
  {https://doi.org/10.1038/nature08812} {\bibfield  {journal} {\bibinfo
  {journal} {Nature}\ }\textbf {\bibinfo {volume} {464}},\ \bibinfo {pages}
  {45} (\bibinfo {year} {2010})}\BibitemShut {NoStop}%
\bibitem [{\citenamefont {Nielsen}\ and\ \citenamefont
  {Chuang}(2010)}]{Nielsen2010}%
  \BibitemOpen
  \bibfield  {author} {\bibinfo {author} {\bibfnamefont {M.~A.}\ \bibnamefont
  {Nielsen}}\ and\ \bibinfo {author} {\bibfnamefont {I.~L.}\ \bibnamefont
  {Chuang}},\ }\href@noop {} {\emph {\bibinfo {title} {Quantum Computation and
  Quantum Information: 10th Anniversary Edition}}}\ (\bibinfo  {publisher}
  {Cambridge University Press},\ \bibinfo {year} {2010})\BibitemShut {NoStop}%
\bibitem [{\citenamefont {Nayak}\ \emph {et~al.}(2008)\citenamefont {Nayak},
  \citenamefont {Simon}, \citenamefont {Stern}, \citenamefont {Freedman},\ and\
  \citenamefont {Das~Sarma}}]{Nayak2008}%
  \BibitemOpen
  \bibfield  {author} {\bibinfo {author} {\bibfnamefont {C.}~\bibnamefont
  {Nayak}}, \bibinfo {author} {\bibfnamefont {S.~H.}\ \bibnamefont {Simon}},
  \bibinfo {author} {\bibfnamefont {A.}~\bibnamefont {Stern}}, \bibinfo
  {author} {\bibfnamefont {M.}~\bibnamefont {Freedman}},\ and\ \bibinfo
  {author} {\bibfnamefont {S.}~\bibnamefont {Das~Sarma}},\ }\bibfield  {title}
  {\bibinfo {title} {Non-abelian anyons and topological quantum computation},\
  }\href {https://doi.org/10.1103/RevModPhys.80.1083} {\bibfield  {journal}
  {\bibinfo  {journal} {Rev. Mod. Phys.}\ }\textbf {\bibinfo {volume} {80}},\
  \bibinfo {pages} {1083} (\bibinfo {year} {2008})}\BibitemShut {NoStop}%
\bibitem [{\citenamefont {Ruby}\ \emph {et~al.}(2017)\citenamefont {Ruby},
  \citenamefont {Heinrich}, \citenamefont {Peng}, \citenamefont {von Oppen},\
  and\ \citenamefont {Franke}}]{KT}%
  \BibitemOpen
  \bibfield  {author} {\bibinfo {author} {\bibfnamefont {M.}~\bibnamefont
  {Ruby}}, \bibinfo {author} {\bibfnamefont {B.~W.}\ \bibnamefont {Heinrich}},
  \bibinfo {author} {\bibfnamefont {Y.}~\bibnamefont {Peng}}, \bibinfo {author}
  {\bibfnamefont {F.}~\bibnamefont {von Oppen}},\ and\ \bibinfo {author}
  {\bibfnamefont {K.~J.}\ \bibnamefont {Franke}},\ }\bibfield  {title}
  {\bibinfo {title} {{Exploring a Proximity-Coupled Co Chain on Pb(110) as a
  Possible Majorana Platform}},\ }\href
  {https://doi.org/10.1021/acs.nanolett.7b01728} {\bibfield  {journal}
  {\bibinfo  {journal} {Nano Letters}\ }\textbf {\bibinfo {volume} {17}},\
  \bibinfo {pages} {4473} (\bibinfo {year} {2017})},\ \bibinfo {note} {pMID:
  28640633},\ \Eprint
  {https://arxiv.org/abs/https://doi.org/10.1021/acs.nanolett.7b01728}
  {https://doi.org/10.1021/acs.nanolett.7b01728} \BibitemShut {NoStop}%
\bibitem [{\citenamefont {Mourik}\ \emph {et~al.}(2012)\citenamefont {Mourik},
  \citenamefont {Zuo}, \citenamefont {Frolov}, \citenamefont {Plissard},
  \citenamefont {Bakkers},\ and\ \citenamefont
  {Kouwenhoven}}]{doi:10.1126/science.1222360}%
  \BibitemOpen
  \bibfield  {author} {\bibinfo {author} {\bibfnamefont {V.}~\bibnamefont
  {Mourik}}, \bibinfo {author} {\bibfnamefont {K.}~\bibnamefont {Zuo}},
  \bibinfo {author} {\bibfnamefont {S.~M.}\ \bibnamefont {Frolov}}, \bibinfo
  {author} {\bibfnamefont {S.~R.}\ \bibnamefont {Plissard}}, \bibinfo {author}
  {\bibfnamefont {E.~P. A.~M.}\ \bibnamefont {Bakkers}},\ and\ \bibinfo
  {author} {\bibfnamefont {L.~P.}\ \bibnamefont {Kouwenhoven}},\ }\bibfield
  {title} {\bibinfo {title} {Signatures of majorana fermions in hybrid
  superconductor-semiconductor nanowire devices},\ }\href
  {https://doi.org/10.1126/science.1222360} {\bibfield  {journal} {\bibinfo
  {journal} {Science}\ }\textbf {\bibinfo {volume} {336}},\ \bibinfo {pages}
  {1003} (\bibinfo {year} {2012})},\ \Eprint
  {https://arxiv.org/abs/https://www.science.org/doi/pdf/10.1126/science.1222360}
  {https://www.science.org/doi/pdf/10.1126/science.1222360} \BibitemShut
  {NoStop}%
\bibitem [{\citenamefont {Nadj-Perge}\ \emph {et~al.}(2014)\citenamefont
  {Nadj-Perge}, \citenamefont {Drozdov}, \citenamefont {Li}, \citenamefont
  {Chen}, \citenamefont {Jeon}, \citenamefont {Seo}, \citenamefont {MacDonald},
  \citenamefont {Bernevig},\ and\ \citenamefont {Yazdani}}]{Nadj2014}%
  \BibitemOpen
  \bibfield  {author} {\bibinfo {author} {\bibfnamefont {S.}~\bibnamefont
  {Nadj-Perge}}, \bibinfo {author} {\bibfnamefont {I.~K.}\ \bibnamefont
  {Drozdov}}, \bibinfo {author} {\bibfnamefont {J.}~\bibnamefont {Li}},
  \bibinfo {author} {\bibfnamefont {H.}~\bibnamefont {Chen}}, \bibinfo {author}
  {\bibfnamefont {S.}~\bibnamefont {Jeon}}, \bibinfo {author} {\bibfnamefont
  {J.}~\bibnamefont {Seo}}, \bibinfo {author} {\bibfnamefont {A.~H.}\
  \bibnamefont {MacDonald}}, \bibinfo {author} {\bibfnamefont {B.~A.}\
  \bibnamefont {Bernevig}},\ and\ \bibinfo {author} {\bibfnamefont
  {A.}~\bibnamefont {Yazdani}},\ }\bibfield  {title} {\bibinfo {title}
  {Observation of majorana fermions in ferromagnetic atomic chains on a
  superconductor},\ }\href {https://doi.org/10.1126/science.1259327} {\bibfield
   {journal} {\bibinfo  {journal} {Science}\ }\textbf {\bibinfo {volume}
  {346}},\ \bibinfo {pages} {602} (\bibinfo {year} {2014})},\ \Eprint
  {https://arxiv.org/abs/https://science.sciencemag.org/content/346/6209/602.full.pdf}
  {https://science.sciencemag.org/content/346/6209/602.full.pdf} \BibitemShut
  {NoStop}%
\bibitem [{\citenamefont {Zhang}\ \emph {et~al.}(2018)\citenamefont {Zhang},
  \citenamefont {Liu}, \citenamefont {Gazibegovic}, \citenamefont {Xu},
  \citenamefont {Logan}, \citenamefont {Wang}, \citenamefont {van Loo},
  \citenamefont {Bommer}, \citenamefont {de~Moor}, \citenamefont {Car},
  \citenamefont {Op~het Veld}, \citenamefont {van Veldhoven}, \citenamefont
  {Koelling}, \citenamefont {Verheijen}, \citenamefont {Pendharkar},
  \citenamefont {Pennachio}, \citenamefont {Shojaei}, \citenamefont {Lee},
  \citenamefont {Palmstrøm}, \citenamefont {Bakkers}, \citenamefont {Sarma},\
  and\ \citenamefont {Kouwenhoven}}]{kouwmajo}%
  \BibitemOpen
  \bibfield  {author} {\bibinfo {author} {\bibfnamefont {H.}~\bibnamefont
  {Zhang}}, \bibinfo {author} {\bibfnamefont {C.-X.}\ \bibnamefont {Liu}},
  \bibinfo {author} {\bibfnamefont {S.}~\bibnamefont {Gazibegovic}}, \bibinfo
  {author} {\bibfnamefont {D.}~\bibnamefont {Xu}}, \bibinfo {author}
  {\bibfnamefont {J.~A.}\ \bibnamefont {Logan}}, \bibinfo {author}
  {\bibfnamefont {G.}~\bibnamefont {Wang}}, \bibinfo {author} {\bibfnamefont
  {N.}~\bibnamefont {van Loo}}, \bibinfo {author} {\bibfnamefont {J.~D.~S.}\
  \bibnamefont {Bommer}}, \bibinfo {author} {\bibfnamefont {M.~W.~A.}\
  \bibnamefont {de~Moor}}, \bibinfo {author} {\bibfnamefont {D.}~\bibnamefont
  {Car}}, \bibinfo {author} {\bibfnamefont {R.~L.~M.}\ \bibnamefont {Op~het
  Veld}}, \bibinfo {author} {\bibfnamefont {P.~J.}\ \bibnamefont {van
  Veldhoven}}, \bibinfo {author} {\bibfnamefont {S.}~\bibnamefont {Koelling}},
  \bibinfo {author} {\bibfnamefont {M.~A.}\ \bibnamefont {Verheijen}}, \bibinfo
  {author} {\bibfnamefont {M.}~\bibnamefont {Pendharkar}}, \bibinfo {author}
  {\bibfnamefont {D.~J.}\ \bibnamefont {Pennachio}}, \bibinfo {author}
  {\bibfnamefont {B.}~\bibnamefont {Shojaei}}, \bibinfo {author} {\bibfnamefont
  {J.~S.}\ \bibnamefont {Lee}}, \bibinfo {author} {\bibfnamefont {C.~J.}\
  \bibnamefont {Palmstrøm}}, \bibinfo {author} {\bibfnamefont {E.~P. A.~M.}\
  \bibnamefont {Bakkers}}, \bibinfo {author} {\bibfnamefont {S.~D.}\
  \bibnamefont {Sarma}},\ and\ \bibinfo {author} {\bibfnamefont {L.~P.}\
  \bibnamefont {Kouwenhoven}},\ }\bibfield  {title} {\bibinfo {title}
  {Quantized majorana conductance},\ }\href@noop {} {\bibfield  {journal}
  {\bibinfo  {journal} {Nature}\ }\textbf {\bibinfo {volume} {556}},\ \bibinfo
  {pages} {74} (\bibinfo {year} {2018})}\BibitemShut {NoStop}%
\bibitem [{\citenamefont {Das}\ \emph {et~al.}(2012)\citenamefont {Das},
  \citenamefont {Ronen}, \citenamefont {Most}, \citenamefont {Oreg},
  \citenamefont {Heiblum},\ and\ \citenamefont {Shtrikman}}]{hbmajo}%
  \BibitemOpen
  \bibfield  {author} {\bibinfo {author} {\bibfnamefont {A.}~\bibnamefont
  {Das}}, \bibinfo {author} {\bibfnamefont {Y.}~\bibnamefont {Ronen}}, \bibinfo
  {author} {\bibfnamefont {Y.}~\bibnamefont {Most}}, \bibinfo {author}
  {\bibfnamefont {Y.}~\bibnamefont {Oreg}}, \bibinfo {author} {\bibfnamefont
  {M.}~\bibnamefont {Heiblum}},\ and\ \bibinfo {author} {\bibfnamefont
  {H.}~\bibnamefont {Shtrikman}},\ }\bibfield  {title} {\bibinfo {title}
  {Zero-bias peaks and splitting in an al–inas nanowire topological
  superconductor as a signature of majorana fermions},\ }\href@noop {}
  {\bibfield  {journal} {\bibinfo  {journal} {Nature Phys.}\ }\textbf {\bibinfo
  {volume} {8}},\ \bibinfo {pages} {887} (\bibinfo {year} {2012})}\BibitemShut
  {NoStop}%
\bibitem [{\citenamefont {Hassler}\ and\ \citenamefont
  {Schuricht}(2012)}]{Hassler_2012}%
  \BibitemOpen
  \bibfield  {author} {\bibinfo {author} {\bibfnamefont {F.}~\bibnamefont
  {Hassler}}\ and\ \bibinfo {author} {\bibfnamefont {D.}~\bibnamefont
  {Schuricht}},\ }\bibfield  {title} {\bibinfo {title} {Strongly interacting
  majorana modes in an array of josephson junctions},\ }\href
  {https://doi.org/10.1088/1367-2630/14/12/125018} {\bibfield  {journal}
  {\bibinfo  {journal} {New Journal of Physics}\ }\textbf {\bibinfo {volume}
  {14}},\ \bibinfo {pages} {125018} (\bibinfo {year} {2012})}\BibitemShut
  {NoStop}%
\bibitem [{\citenamefont {Sela}\ \emph {et~al.}(2011)\citenamefont {Sela},
  \citenamefont {Altland},\ and\ \citenamefont {Rosch}}]{PhysRevB.84.085114}%
  \BibitemOpen
  \bibfield  {author} {\bibinfo {author} {\bibfnamefont {E.}~\bibnamefont
  {Sela}}, \bibinfo {author} {\bibfnamefont {A.}~\bibnamefont {Altland}},\ and\
  \bibinfo {author} {\bibfnamefont {A.}~\bibnamefont {Rosch}},\ }\bibfield
  {title} {\bibinfo {title} {Majorana fermions in strongly interacting helical
  liquids},\ }\href {https://doi.org/10.1103/PhysRevB.84.085114} {\bibfield
  {journal} {\bibinfo  {journal} {Phys. Rev. B}\ }\textbf {\bibinfo {volume}
  {84}},\ \bibinfo {pages} {085114} (\bibinfo {year} {2011})}\BibitemShut
  {NoStop}%
\bibitem [{\citenamefont {Thomale}\ \emph {et~al.}(2013)\citenamefont
  {Thomale}, \citenamefont {Rachel},\ and\ \citenamefont
  {Schmitteckert}}]{PhysRevB.88.161103}%
  \BibitemOpen
  \bibfield  {author} {\bibinfo {author} {\bibfnamefont {R.}~\bibnamefont
  {Thomale}}, \bibinfo {author} {\bibfnamefont {S.}~\bibnamefont {Rachel}},\
  and\ \bibinfo {author} {\bibfnamefont {P.}~\bibnamefont {Schmitteckert}},\
  }\bibfield  {title} {\bibinfo {title} {Tunneling spectra simulation of
  interacting majorana wires},\ }\href
  {https://doi.org/10.1103/PhysRevB.88.161103} {\bibfield  {journal} {\bibinfo
  {journal} {Phys. Rev. B}\ }\textbf {\bibinfo {volume} {88}},\ \bibinfo
  {pages} {161103} (\bibinfo {year} {2013})}\BibitemShut {NoStop}%
\bibitem [{\citenamefont {Katsura}\ \emph {et~al.}(2015)\citenamefont
  {Katsura}, \citenamefont {Schuricht},\ and\ \citenamefont
  {Takahashi}}]{Katsura2015}%
  \BibitemOpen
  \bibfield  {author} {\bibinfo {author} {\bibfnamefont {H.}~\bibnamefont
  {Katsura}}, \bibinfo {author} {\bibfnamefont {D.}~\bibnamefont {Schuricht}},\
  and\ \bibinfo {author} {\bibfnamefont {M.}~\bibnamefont {Takahashi}},\
  }\bibfield  {title} {\bibinfo {title} {Exact ground states and topological
  order in interacting kitaev/majorana chains},\ }\href
  {https://doi.org/10.1103/PhysRevB.92.115137} {\bibfield  {journal} {\bibinfo
  {journal} {Phys. Rev. B}\ }\textbf {\bibinfo {volume} {92}},\ \bibinfo
  {pages} {115137} (\bibinfo {year} {2015})}\BibitemShut {NoStop}%
\bibitem [{\citenamefont {Mahyaeh}\ and\ \citenamefont
  {Ardonne}(2020)}]{Mahyaeh2020}%
  \BibitemOpen
  \bibfield  {author} {\bibinfo {author} {\bibfnamefont {I.}~\bibnamefont
  {Mahyaeh}}\ and\ \bibinfo {author} {\bibfnamefont {E.}~\bibnamefont
  {Ardonne}},\ }\bibfield  {title} {\bibinfo {title} {Study of the phase
  diagram of the kitaev-hubbard chain},\ }\href
  {https://doi.org/10.1103/PhysRevB.101.085125} {\bibfield  {journal} {\bibinfo
   {journal} {Phys. Rev. B}\ }\textbf {\bibinfo {volume} {101}},\ \bibinfo
  {pages} {085125} (\bibinfo {year} {2020})}\BibitemShut {NoStop}%
\bibitem [{\citenamefont {Miao}\ \emph {et~al.}(2017)\citenamefont {Miao},
  \citenamefont {Jin}, \citenamefont {Zhang},\ and\ \citenamefont
  {Zhou}}]{PhysRevLett.118.267701}%
  \BibitemOpen
  \bibfield  {author} {\bibinfo {author} {\bibfnamefont {J.-J.}\ \bibnamefont
  {Miao}}, \bibinfo {author} {\bibfnamefont {H.-K.}\ \bibnamefont {Jin}},
  \bibinfo {author} {\bibfnamefont {F.-C.}\ \bibnamefont {Zhang}},\ and\
  \bibinfo {author} {\bibfnamefont {Y.}~\bibnamefont {Zhou}},\ }\bibfield
  {title} {\bibinfo {title} {Exact solution for the interacting kitaev chain at
  the symmetric point},\ }\href
  {https://doi.org/10.1103/PhysRevLett.118.267701} {\bibfield  {journal}
  {\bibinfo  {journal} {Phys. Rev. Lett.}\ }\textbf {\bibinfo {volume} {118}},\
  \bibinfo {pages} {267701} (\bibinfo {year} {2017})}\BibitemShut {NoStop}%
\bibitem [{\citenamefont {Chan}\ \emph {et~al.}(2015)\citenamefont {Chan},
  \citenamefont {Chiu},\ and\ \citenamefont {Sun}}]{PhysRevB.92.104514}%
  \BibitemOpen
  \bibfield  {author} {\bibinfo {author} {\bibfnamefont {Y.-H.}\ \bibnamefont
  {Chan}}, \bibinfo {author} {\bibfnamefont {C.-K.}\ \bibnamefont {Chiu}},\
  and\ \bibinfo {author} {\bibfnamefont {K.}~\bibnamefont {Sun}},\ }\bibfield
  {title} {\bibinfo {title} {Multiple signatures of topological transitions for
  interacting fermions in chain lattices},\ }\href
  {https://doi.org/10.1103/PhysRevB.92.104514} {\bibfield  {journal} {\bibinfo
  {journal} {Phys. Rev. B}\ }\textbf {\bibinfo {volume} {92}},\ \bibinfo
  {pages} {104514} (\bibinfo {year} {2015})}\BibitemShut {NoStop}%
\bibitem [{\citenamefont {DeGottardi}\ \emph
  {et~al.}(2013{\natexlab{a}})\citenamefont {DeGottardi}, \citenamefont {Sen},\
  and\ \citenamefont {Vishveshwara}}]{DeGottardi2013}%
  \BibitemOpen
  \bibfield  {author} {\bibinfo {author} {\bibfnamefont {W.}~\bibnamefont
  {DeGottardi}}, \bibinfo {author} {\bibfnamefont {D.}~\bibnamefont {Sen}},\
  and\ \bibinfo {author} {\bibfnamefont {S.}~\bibnamefont {Vishveshwara}},\
  }\bibfield  {title} {\bibinfo {title} {Majorana fermions in superconducting
  1d systems having periodic, quasiperiodic, and disordered potentials},\
  }\href {https://doi.org/10.1103/PhysRevLett.110.146404} {\bibfield  {journal}
  {\bibinfo  {journal} {Phys. Rev. Lett.}\ }\textbf {\bibinfo {volume} {110}},\
  \bibinfo {pages} {146404} (\bibinfo {year} {2013}{\natexlab{a}})}\BibitemShut
  {NoStop}%
\bibitem [{\citenamefont {DeGottardi}\ \emph
  {et~al.}(2013{\natexlab{b}})\citenamefont {DeGottardi}, \citenamefont
  {Thakurathi}, \citenamefont {Vishveshwara},\ and\ \citenamefont
  {Sen}}]{DeGottardi2013b}%
  \BibitemOpen
  \bibfield  {author} {\bibinfo {author} {\bibfnamefont {W.}~\bibnamefont
  {DeGottardi}}, \bibinfo {author} {\bibfnamefont {M.}~\bibnamefont
  {Thakurathi}}, \bibinfo {author} {\bibfnamefont {S.}~\bibnamefont
  {Vishveshwara}},\ and\ \bibinfo {author} {\bibfnamefont {D.}~\bibnamefont
  {Sen}},\ }\bibfield  {title} {\bibinfo {title} {Majorana fermions in
  superconducting wires: Effects of long-range hopping, broken time-reversal
  symmetry, and potential landscapes},\ }\href
  {https://doi.org/10.1103/PhysRevB.88.165111} {\bibfield  {journal} {\bibinfo
  {journal} {Phys. Rev. B}\ }\textbf {\bibinfo {volume} {88}},\ \bibinfo
  {pages} {165111} (\bibinfo {year} {2013}{\natexlab{b}})}\BibitemShut
  {NoStop}%
\bibitem [{\citenamefont {Altland}\ \emph {et~al.}(2014)\citenamefont
  {Altland}, \citenamefont {Bagrets}, \citenamefont {Fritz}, \citenamefont
  {Kamenev},\ and\ \citenamefont {Schmiedt}}]{Altland2014}%
  \BibitemOpen
  \bibfield  {author} {\bibinfo {author} {\bibfnamefont {A.}~\bibnamefont
  {Altland}}, \bibinfo {author} {\bibfnamefont {D.}~\bibnamefont {Bagrets}},
  \bibinfo {author} {\bibfnamefont {L.}~\bibnamefont {Fritz}}, \bibinfo
  {author} {\bibfnamefont {A.}~\bibnamefont {Kamenev}},\ and\ \bibinfo {author}
  {\bibfnamefont {H.}~\bibnamefont {Schmiedt}},\ }\bibfield  {title} {\bibinfo
  {title} {Quantum criticality of quasi-one-dimensional topological anderson
  insulators},\ }\href {https://doi.org/10.1103/PhysRevLett.112.206602}
  {\bibfield  {journal} {\bibinfo  {journal} {Phys. Rev. Lett.}\ }\textbf
  {\bibinfo {volume} {112}},\ \bibinfo {pages} {206602} (\bibinfo {year}
  {2014})}\BibitemShut {NoStop}%
\bibitem [{\citenamefont {Francica}\ \emph {et~al.}(2023)\citenamefont
  {Francica}, \citenamefont {Tiburzi},\ and\ \citenamefont
  {Dell'Anna}}]{PhysRevB.107.165137}%
  \BibitemOpen
  \bibfield  {author} {\bibinfo {author} {\bibfnamefont {G.}~\bibnamefont
  {Francica}}, \bibinfo {author} {\bibfnamefont {E.~M.}\ \bibnamefont
  {Tiburzi}},\ and\ \bibinfo {author} {\bibfnamefont {L.}~\bibnamefont
  {Dell'Anna}},\ }\bibfield  {title} {\bibinfo {title} {Topological phases in
  the presence of disorder and longer-range couplings},\ }\href
  {https://doi.org/10.1103/PhysRevB.107.165137} {\bibfield  {journal} {\bibinfo
   {journal} {Phys. Rev. B}\ }\textbf {\bibinfo {volume} {107}},\ \bibinfo
  {pages} {165137} (\bibinfo {year} {2023})}\BibitemShut {NoStop}%
\bibitem [{\citenamefont {Lobos}\ \emph {et~al.}(2012)\citenamefont {Lobos},
  \citenamefont {Lutchyn},\ and\ \citenamefont {Das~Sarma}}]{Lobos2012}%
  \BibitemOpen
  \bibfield  {author} {\bibinfo {author} {\bibfnamefont {A.~M.}\ \bibnamefont
  {Lobos}}, \bibinfo {author} {\bibfnamefont {R.~M.}\ \bibnamefont {Lutchyn}},\
  and\ \bibinfo {author} {\bibfnamefont {S.}~\bibnamefont {Das~Sarma}},\
  }\bibfield  {title} {\bibinfo {title} {Interplay of disorder and interaction
  in majorana quantum wires},\ }\href
  {https://doi.org/10.1103/PhysRevLett.109.146403} {\bibfield  {journal}
  {\bibinfo  {journal} {Phys. Rev. Lett.}\ }\textbf {\bibinfo {volume} {109}},\
  \bibinfo {pages} {146403} (\bibinfo {year} {2012})}\BibitemShut {NoStop}%
\bibitem [{\citenamefont {Cr\'epin}\ \emph {et~al.}(2014)\citenamefont
  {Cr\'epin}, \citenamefont {Zar\'and},\ and\ \citenamefont
  {Simon}}]{Crepin2014}%
  \BibitemOpen
  \bibfield  {author} {\bibinfo {author} {\bibfnamefont {F.~m.~c.}\
  \bibnamefont {Cr\'epin}}, \bibinfo {author} {\bibfnamefont {G.}~\bibnamefont
  {Zar\'and}},\ and\ \bibinfo {author} {\bibfnamefont {P.}~\bibnamefont
  {Simon}},\ }\bibfield  {title} {\bibinfo {title} {Nonperturbative phase
  diagram of interacting disordered majorana nanowires},\ }\href
  {https://doi.org/10.1103/PhysRevB.90.121407} {\bibfield  {journal} {\bibinfo
  {journal} {Phys. Rev. B}\ }\textbf {\bibinfo {volume} {90}},\ \bibinfo
  {pages} {121407} (\bibinfo {year} {2014})}\BibitemShut {NoStop}%
\bibitem [{\citenamefont {Gangadharaiah}\ \emph {et~al.}(2011)\citenamefont
  {Gangadharaiah}, \citenamefont {Braunecker}, \citenamefont {Simon},\ and\
  \citenamefont {Loss}}]{PhysRevLett.107.036801}%
  \BibitemOpen
  \bibfield  {author} {\bibinfo {author} {\bibfnamefont {S.}~\bibnamefont
  {Gangadharaiah}}, \bibinfo {author} {\bibfnamefont {B.}~\bibnamefont
  {Braunecker}}, \bibinfo {author} {\bibfnamefont {P.}~\bibnamefont {Simon}},\
  and\ \bibinfo {author} {\bibfnamefont {D.}~\bibnamefont {Loss}},\ }\bibfield
  {title} {\bibinfo {title} {Majorana edge states in interacting
  one-dimensional systems},\ }\href
  {https://doi.org/10.1103/PhysRevLett.107.036801} {\bibfield  {journal}
  {\bibinfo  {journal} {Phys. Rev. Lett.}\ }\textbf {\bibinfo {volume} {107}},\
  \bibinfo {pages} {036801} (\bibinfo {year} {2011})}\BibitemShut {NoStop}%
\bibitem [{\citenamefont {Stoudenmire}\ \emph {et~al.}(2011)\citenamefont
  {Stoudenmire}, \citenamefont {Alicea}, \citenamefont {Starykh},\ and\
  \citenamefont {Fisher}}]{PhysRevB.84.014503}%
  \BibitemOpen
  \bibfield  {author} {\bibinfo {author} {\bibfnamefont {E.~M.}\ \bibnamefont
  {Stoudenmire}}, \bibinfo {author} {\bibfnamefont {J.}~\bibnamefont {Alicea}},
  \bibinfo {author} {\bibfnamefont {O.~A.}\ \bibnamefont {Starykh}},\ and\
  \bibinfo {author} {\bibfnamefont {M.~P.}\ \bibnamefont {Fisher}},\ }\bibfield
   {title} {\bibinfo {title} {Interaction effects in topological
  superconducting wires supporting majorana fermions},\ }\href
  {https://doi.org/10.1103/PhysRevB.84.014503} {\bibfield  {journal} {\bibinfo
  {journal} {Phys. Rev. B}\ }\textbf {\bibinfo {volume} {84}},\ \bibinfo
  {pages} {014503} (\bibinfo {year} {2011})}\BibitemShut {NoStop}%
\bibitem [{\citenamefont {Cheng}\ and\ \citenamefont
  {Tu}(2011)}]{PhysRevB.84.094503}%
  \BibitemOpen
  \bibfield  {author} {\bibinfo {author} {\bibfnamefont {M.}~\bibnamefont
  {Cheng}}\ and\ \bibinfo {author} {\bibfnamefont {H.-H.}\ \bibnamefont {Tu}},\
  }\bibfield  {title} {\bibinfo {title} {Majorana edge states in interacting
  two-chain ladders of fermions},\ }\href
  {https://doi.org/10.1103/PhysRevB.84.094503} {\bibfield  {journal} {\bibinfo
  {journal} {Phys. Rev. B}\ }\textbf {\bibinfo {volume} {84}},\ \bibinfo
  {pages} {094503} (\bibinfo {year} {2011})}\BibitemShut {NoStop}%
\bibitem [{\citenamefont {Lutchyn}\ and\ \citenamefont
  {Fisher}(2011)}]{PhysRevB.84.214528}%
  \BibitemOpen
  \bibfield  {author} {\bibinfo {author} {\bibfnamefont {R.~M.}\ \bibnamefont
  {Lutchyn}}\ and\ \bibinfo {author} {\bibfnamefont {M.~P.~A.}\ \bibnamefont
  {Fisher}},\ }\bibfield  {title} {\bibinfo {title} {Interacting topological
  phases in multiband nanowires},\ }\href
  {https://doi.org/10.1103/PhysRevB.84.214528} {\bibfield  {journal} {\bibinfo
  {journal} {Phys. Rev. B}\ }\textbf {\bibinfo {volume} {84}},\ \bibinfo
  {pages} {214528} (\bibinfo {year} {2011})}\BibitemShut {NoStop}%
\bibitem [{\citenamefont {Manolescu}\ \emph {et~al.}(2014)\citenamefont
  {Manolescu}, \citenamefont {Marinescu},\ and\ \citenamefont
  {Stanescu}}]{Manolescu2014}%
  \BibitemOpen
  \bibfield  {author} {\bibinfo {author} {\bibfnamefont {A.}~\bibnamefont
  {Manolescu}}, \bibinfo {author} {\bibfnamefont {D.~C.}\ \bibnamefont
  {Marinescu}},\ and\ \bibinfo {author} {\bibfnamefont {T.~D.}\ \bibnamefont
  {Stanescu}},\ }\bibfield  {title} {\bibinfo {title} {Coulomb interaction
  effects on the majorana states in quantum wires},\ }\href
  {https://doi.org/10.1088/0953-8984/26/17/172203} {\bibfield  {journal}
  {\bibinfo  {journal} {J. Phys.: Condens. Matter}\ }\textbf {\bibinfo {volume}
  {26}},\ \bibinfo {pages} {172203} (\bibinfo {year} {2014})}\BibitemShut
  {NoStop}%
\bibitem [{\citenamefont {Rahmani}\ \emph
  {et~al.}(2015{\natexlab{a}})\citenamefont {Rahmani}, \citenamefont {Zhu},
  \citenamefont {Franz},\ and\ \citenamefont
  {Affleck}}]{PhysRevLett.115.166401}%
  \BibitemOpen
  \bibfield  {author} {\bibinfo {author} {\bibfnamefont {A.}~\bibnamefont
  {Rahmani}}, \bibinfo {author} {\bibfnamefont {X.}~\bibnamefont {Zhu}},
  \bibinfo {author} {\bibfnamefont {M.}~\bibnamefont {Franz}},\ and\ \bibinfo
  {author} {\bibfnamefont {I.}~\bibnamefont {Affleck}},\ }\bibfield  {title}
  {\bibinfo {title} {Emergent supersymmetry from strongly interacting majorana
  zero modes},\ }\href {https://doi.org/10.1103/PhysRevLett.115.166401}
  {\bibfield  {journal} {\bibinfo  {journal} {Phys. Rev. Lett.}\ }\textbf
  {\bibinfo {volume} {115}},\ \bibinfo {pages} {166401} (\bibinfo {year}
  {2015}{\natexlab{a}})}\BibitemShut {NoStop}%
\bibitem [{\citenamefont {Kells}(2015{\natexlab{a}})}]{PhysRevB.92.081401}%
  \BibitemOpen
  \bibfield  {author} {\bibinfo {author} {\bibfnamefont {G.}~\bibnamefont
  {Kells}},\ }\bibfield  {title} {\bibinfo {title} {Many-body majorana
  operators and the equivalence of parity sectors},\ }\href
  {https://doi.org/10.1103/PhysRevB.92.081401} {\bibfield  {journal} {\bibinfo
  {journal} {Phys. Rev. B}\ }\textbf {\bibinfo {volume} {92}},\ \bibinfo
  {pages} {081401} (\bibinfo {year} {2015}{\natexlab{a}})}\BibitemShut
  {NoStop}%
\bibitem [{\citenamefont {Milsted}\ \emph {et~al.}(2015)\citenamefont
  {Milsted}, \citenamefont {Seabra}, \citenamefont {Fulga}, \citenamefont
  {Beenakker},\ and\ \citenamefont {Cobanera}}]{PhysRevB.92.085139}%
  \BibitemOpen
  \bibfield  {author} {\bibinfo {author} {\bibfnamefont {A.}~\bibnamefont
  {Milsted}}, \bibinfo {author} {\bibfnamefont {L.}~\bibnamefont {Seabra}},
  \bibinfo {author} {\bibfnamefont {I.~C.}\ \bibnamefont {Fulga}}, \bibinfo
  {author} {\bibfnamefont {C.~W.~J.}\ \bibnamefont {Beenakker}},\ and\ \bibinfo
  {author} {\bibfnamefont {E.}~\bibnamefont {Cobanera}},\ }\bibfield  {title}
  {\bibinfo {title} {Statistical translation invariance protects a topological
  insulator from interactions},\ }\href
  {https://doi.org/10.1103/PhysRevB.92.085139} {\bibfield  {journal} {\bibinfo
  {journal} {Phys. Rev. B}\ }\textbf {\bibinfo {volume} {92}},\ \bibinfo
  {pages} {085139} (\bibinfo {year} {2015})}\BibitemShut {NoStop}%
\bibitem [{\citenamefont {Kells}(2015{\natexlab{b}})}]{PhysRevB.92.155434}%
  \BibitemOpen
  \bibfield  {author} {\bibinfo {author} {\bibfnamefont {G.}~\bibnamefont
  {Kells}},\ }\bibfield  {title} {\bibinfo {title} {Multiparticle content of
  majorana zero modes in the interacting $p$-wave wire},\ }\href
  {https://doi.org/10.1103/PhysRevB.92.155434} {\bibfield  {journal} {\bibinfo
  {journal} {Phys. Rev. B}\ }\textbf {\bibinfo {volume} {92}},\ \bibinfo
  {pages} {155434} (\bibinfo {year} {2015}{\natexlab{b}})}\BibitemShut
  {NoStop}%
\bibitem [{\citenamefont {Rahmani}\ \emph
  {et~al.}(2015{\natexlab{b}})\citenamefont {Rahmani}, \citenamefont {Zhu},
  \citenamefont {Franz},\ and\ \citenamefont {Affleck}}]{PhysRevB.92.235123}%
  \BibitemOpen
  \bibfield  {author} {\bibinfo {author} {\bibfnamefont {A.}~\bibnamefont
  {Rahmani}}, \bibinfo {author} {\bibfnamefont {X.}~\bibnamefont {Zhu}},
  \bibinfo {author} {\bibfnamefont {M.}~\bibnamefont {Franz}},\ and\ \bibinfo
  {author} {\bibfnamefont {I.}~\bibnamefont {Affleck}},\ }\bibfield  {title}
  {\bibinfo {title} {Phase diagram of the interacting majorana chain model},\
  }\href {https://doi.org/10.1103/PhysRevB.92.235123} {\bibfield  {journal}
  {\bibinfo  {journal} {Phys. Rev. B}\ }\textbf {\bibinfo {volume} {92}},\
  \bibinfo {pages} {235123} (\bibinfo {year} {2015}{\natexlab{b}})}\BibitemShut
  {NoStop}%
\bibitem [{\citenamefont {Brouwer}\ \emph {et~al.}(2011)\citenamefont
  {Brouwer}, \citenamefont {Duckheim}, \citenamefont {Romito},\ and\
  \citenamefont {von Oppen}}]{Brouwer2011}%
  \BibitemOpen
  \bibfield  {author} {\bibinfo {author} {\bibfnamefont {P.~W.}\ \bibnamefont
  {Brouwer}}, \bibinfo {author} {\bibfnamefont {M.}~\bibnamefont {Duckheim}},
  \bibinfo {author} {\bibfnamefont {A.}~\bibnamefont {Romito}},\ and\ \bibinfo
  {author} {\bibfnamefont {F.}~\bibnamefont {von Oppen}},\ }\bibfield  {title}
  {\bibinfo {title} {Probability distribution of majorana end-state energies in
  disordered wires},\ }\href {https://doi.org/10.1103/PhysRevLett.107.196804}
  {\bibfield  {journal} {\bibinfo  {journal} {Phys. Rev. Lett.}\ }\textbf
  {\bibinfo {volume} {107}},\ \bibinfo {pages} {196804} (\bibinfo {year}
  {2011})}\BibitemShut {NoStop}%
\bibitem [{\citenamefont {Pientka}\ \emph {et~al.}(2013)\citenamefont
  {Pientka}, \citenamefont {Romito}, \citenamefont {Duckheim}, \citenamefont
  {Oreg},\ and\ \citenamefont {von Oppen}}]{Pientka2013}%
  \BibitemOpen
  \bibfield  {author} {\bibinfo {author} {\bibfnamefont {F.}~\bibnamefont
  {Pientka}}, \bibinfo {author} {\bibfnamefont {A.}~\bibnamefont {Romito}},
  \bibinfo {author} {\bibfnamefont {M.}~\bibnamefont {Duckheim}}, \bibinfo
  {author} {\bibfnamefont {Y.}~\bibnamefont {Oreg}},\ and\ \bibinfo {author}
  {\bibfnamefont {F.}~\bibnamefont {von Oppen}},\ }\bibfield  {title} {\bibinfo
  {title} {Signatures of topological phase transitions in mesoscopic
  superconducting rings},\ }\href
  {https://doi.org/10.1088/1367-2630/15/2/025001} {\bibfield  {journal}
  {\bibinfo  {journal} {New J. Phys.}\ }\textbf {\bibinfo {volume} {15}},\
  \bibinfo {pages} {025001} (\bibinfo {year} {2013})}\BibitemShut {NoStop}%
\bibitem [{\citenamefont {Lin}\ \emph {et~al.}(2021)\citenamefont {Lin},
  \citenamefont {Weber}, \citenamefont {Kennes}, \citenamefont {Pletyukhov},
  \citenamefont {Schoeller},\ and\ \citenamefont
  {Meden}}]{PhysRevB.103.195119}%
  \BibitemOpen
  \bibfield  {author} {\bibinfo {author} {\bibfnamefont {Y.-T.}\ \bibnamefont
  {Lin}}, \bibinfo {author} {\bibfnamefont {C.~S.}\ \bibnamefont {Weber}},
  \bibinfo {author} {\bibfnamefont {D.~M.}\ \bibnamefont {Kennes}}, \bibinfo
  {author} {\bibfnamefont {M.}~\bibnamefont {Pletyukhov}}, \bibinfo {author}
  {\bibfnamefont {H.}~\bibnamefont {Schoeller}},\ and\ \bibinfo {author}
  {\bibfnamefont {V.}~\bibnamefont {Meden}},\ }\bibfield  {title} {\bibinfo
  {title} {Quantitative analysis of interaction effects in generalized
  aubry-andr\'e-harper models},\ }\href
  {https://doi.org/10.1103/PhysRevB.103.195119} {\bibfield  {journal} {\bibinfo
   {journal} {Phys. Rev. B}\ }\textbf {\bibinfo {volume} {103}},\ \bibinfo
  {pages} {195119} (\bibinfo {year} {2021})}\BibitemShut {NoStop}%
\bibitem [{\citenamefont {Madeira}\ and\ \citenamefont
  {Sacramento}(2022)}]{PhysRevB.106.224505}%
  \BibitemOpen
  \bibfield  {author} {\bibinfo {author} {\bibfnamefont {M.~F.}\ \bibnamefont
  {Madeira}}\ and\ \bibinfo {author} {\bibfnamefont {P.~D.}\ \bibnamefont
  {Sacramento}},\ }\bibfield  {title} {\bibinfo {title} {Quasidisorder-induced
  topology},\ }\href {https://doi.org/10.1103/PhysRevB.106.224505} {\bibfield
  {journal} {\bibinfo  {journal} {Phys. Rev. B}\ }\textbf {\bibinfo {volume}
  {106}},\ \bibinfo {pages} {224505} (\bibinfo {year} {2022})}\BibitemShut
  {NoStop}%
\bibitem [{\citenamefont {Roy}\ \emph {et~al.}(2023)\citenamefont {Roy},
  \citenamefont {Nabi},\ and\ \citenamefont {Basu}}]{PhysRevB.107.014202}%
  \BibitemOpen
  \bibfield  {author} {\bibinfo {author} {\bibfnamefont {S.}~\bibnamefont
  {Roy}}, \bibinfo {author} {\bibfnamefont {S.~N.}\ \bibnamefont {Nabi}},\ and\
  \bibinfo {author} {\bibfnamefont {S.}~\bibnamefont {Basu}},\ }\bibfield
  {title} {\bibinfo {title} {Critical and topological phases of dimerized
  kitaev chain in presence of quasiperiodic potential},\ }\href
  {https://doi.org/10.1103/PhysRevB.107.014202} {\bibfield  {journal} {\bibinfo
   {journal} {Phys. Rev. B}\ }\textbf {\bibinfo {volume} {107}},\ \bibinfo
  {pages} {014202} (\bibinfo {year} {2023})}\BibitemShut {NoStop}%
\bibitem [{\citenamefont {Gon\ifmmode~\mbox{\c{c}}\else \c{c}\fi{}alves}\ \emph
  {et~al.}(2023)\citenamefont {Gon\ifmmode~\mbox{\c{c}}\else \c{c}\fi{}alves},
  \citenamefont {Amorim}, \citenamefont {Castro},\ and\ \citenamefont
  {Ribeiro}}]{PhysRevLett.131.186303}%
  \BibitemOpen
  \bibfield  {author} {\bibinfo {author} {\bibfnamefont {M.}~\bibnamefont
  {Gon\ifmmode~\mbox{\c{c}}\else \c{c}\fi{}alves}}, \bibinfo {author}
  {\bibfnamefont {B.}~\bibnamefont {Amorim}}, \bibinfo {author} {\bibfnamefont
  {E.~V.}\ \bibnamefont {Castro}},\ and\ \bibinfo {author} {\bibfnamefont
  {P.}~\bibnamefont {Ribeiro}},\ }\bibfield  {title} {\bibinfo {title}
  {Critical phase dualities in 1d exactly solvable quasiperiodic models},\
  }\href {https://doi.org/10.1103/PhysRevLett.131.186303} {\bibfield  {journal}
  {\bibinfo  {journal} {Phys. Rev. Lett.}\ }\textbf {\bibinfo {volume} {131}},\
  \bibinfo {pages} {186303} (\bibinfo {year} {2023})}\BibitemShut {NoStop}%
\bibitem [{\citenamefont {Laflorencie}\ \emph {et~al.}(2022)\citenamefont
  {Laflorencie}, \citenamefont {Lemari\'e},\ and\ \citenamefont
  {Mac\'e}}]{PhysRevResearch.4.L032016}%
  \BibitemOpen
  \bibfield  {author} {\bibinfo {author} {\bibfnamefont {N.}~\bibnamefont
  {Laflorencie}}, \bibinfo {author} {\bibfnamefont {G.}~\bibnamefont
  {Lemari\'e}},\ and\ \bibinfo {author} {\bibfnamefont {N.}~\bibnamefont
  {Mac\'e}},\ }\bibfield  {title} {\bibinfo {title} {Topological order in
  random interacting ising-majorana chains stabilized by many-body
  localization},\ }\href {https://doi.org/10.1103/PhysRevResearch.4.L032016}
  {\bibfield  {journal} {\bibinfo  {journal} {Phys. Rev. Res.}\ }\textbf
  {\bibinfo {volume} {4}},\ \bibinfo {pages} {L032016} (\bibinfo {year}
  {2022})}\BibitemShut {NoStop}%
\bibitem [{\citenamefont {Chepiga}\ and\ \citenamefont
  {Laflorencie}(2023)}]{10.21468/SciPostPhys.14.6.152}%
  \BibitemOpen
  \bibfield  {author} {\bibinfo {author} {\bibfnamefont {N.}~\bibnamefont
  {Chepiga}}\ and\ \bibinfo {author} {\bibfnamefont {N.}~\bibnamefont
  {Laflorencie}},\ }\bibfield  {title} {\bibinfo {title} {{Topological and
  quantum critical properties of the interacting Majorana chain model}},\
  }\href {https://doi.org/10.21468/SciPostPhys.14.6.152} {\bibfield  {journal}
  {\bibinfo  {journal} {SciPost Phys.}\ }\textbf {\bibinfo {volume} {14}},\
  \bibinfo {pages} {152} (\bibinfo {year} {2023})}\BibitemShut {NoStop}%
\bibitem [{\citenamefont {Karcher}\ \emph {et~al.}(2019)\citenamefont
  {Karcher}, \citenamefont {Sonner},\ and\ \citenamefont
  {Mirlin}}]{PhysRevB.100.134207}%
  \BibitemOpen
  \bibfield  {author} {\bibinfo {author} {\bibfnamefont {J.~F.}\ \bibnamefont
  {Karcher}}, \bibinfo {author} {\bibfnamefont {M.}~\bibnamefont {Sonner}},\
  and\ \bibinfo {author} {\bibfnamefont {A.~D.}\ \bibnamefont {Mirlin}},\
  }\bibfield  {title} {\bibinfo {title} {Disorder and interaction in chiral
  chains: Majoranas versus complex fermions},\ }\href
  {https://doi.org/10.1103/PhysRevB.100.134207} {\bibfield  {journal} {\bibinfo
   {journal} {Phys. Rev. B}\ }\textbf {\bibinfo {volume} {100}},\ \bibinfo
  {pages} {134207} (\bibinfo {year} {2019})}\BibitemShut {NoStop}%
\bibitem [{\citenamefont {del Pozo}\ \emph {et~al.}(2023)\citenamefont {del
  Pozo}, \citenamefont {Herviou},\ and\ \citenamefont
  {Le~Hur}}]{PhysRevB.107.155134}%
  \BibitemOpen
  \bibfield  {author} {\bibinfo {author} {\bibfnamefont {F.}~\bibnamefont {del
  Pozo}}, \bibinfo {author} {\bibfnamefont {L.}~\bibnamefont {Herviou}},\ and\
  \bibinfo {author} {\bibfnamefont {K.}~\bibnamefont {Le~Hur}},\ }\bibfield
  {title} {\bibinfo {title} {Fractional topology in interacting one-dimensional
  superconductors},\ }\href {https://doi.org/10.1103/PhysRevB.107.155134}
  {\bibfield  {journal} {\bibinfo  {journal} {Phys. Rev. B}\ }\textbf {\bibinfo
  {volume} {107}},\ \bibinfo {pages} {155134} (\bibinfo {year}
  {2023})}\BibitemShut {NoStop}%
\bibitem [{\citenamefont {Shi}\ \emph {et~al.}(2022)\citenamefont {Shi},
  \citenamefont {Zhang},\ and\ \citenamefont {Song}}]{PhysRevB.106.184505}%
  \BibitemOpen
  \bibfield  {author} {\bibinfo {author} {\bibfnamefont {Y.~B.}\ \bibnamefont
  {Shi}}, \bibinfo {author} {\bibfnamefont {K.~L.}\ \bibnamefont {Zhang}},\
  and\ \bibinfo {author} {\bibfnamefont {Z.}~\bibnamefont {Song}},\ }\bibfield
  {title} {\bibinfo {title} {Dynamic generation of nonequilibrium
  superconducting states in a kitaev chain},\ }\href
  {https://doi.org/10.1103/PhysRevB.106.184505} {\bibfield  {journal} {\bibinfo
   {journal} {Phys. Rev. B}\ }\textbf {\bibinfo {volume} {106}},\ \bibinfo
  {pages} {184505} (\bibinfo {year} {2022})}\BibitemShut {NoStop}%
\bibitem [{\citenamefont {Shi}\ and\ \citenamefont
  {Song}(2023)}]{PhysRevB.107.125110}%
  \BibitemOpen
  \bibfield  {author} {\bibinfo {author} {\bibfnamefont {Y.~B.}\ \bibnamefont
  {Shi}}\ and\ \bibinfo {author} {\bibfnamefont {Z.}~\bibnamefont {Song}},\
  }\bibfield  {title} {\bibinfo {title} {Topological phase in a kitaev chain
  with spatially separated pairing processes},\ }\href
  {https://doi.org/10.1103/PhysRevB.107.125110} {\bibfield  {journal} {\bibinfo
   {journal} {Phys. Rev. B}\ }\textbf {\bibinfo {volume} {107}},\ \bibinfo
  {pages} {125110} (\bibinfo {year} {2023})}\BibitemShut {NoStop}%
\bibitem [{\citenamefont {Gergs}\ \emph {et~al.}(2016)\citenamefont {Gergs},
  \citenamefont {Fritz},\ and\ \citenamefont {Schuricht}}]{Gergs2016}%
  \BibitemOpen
  \bibfield  {author} {\bibinfo {author} {\bibfnamefont {N.~M.}\ \bibnamefont
  {Gergs}}, \bibinfo {author} {\bibfnamefont {L.}~\bibnamefont {Fritz}},\ and\
  \bibinfo {author} {\bibfnamefont {D.}~\bibnamefont {Schuricht}},\ }\bibfield
  {title} {\bibinfo {title} {Topological order in the kitaev/majorana chain in
  the presence of disorder and interactions},\ }\href
  {https://doi.org/10.1103/PhysRevB.93.075129} {\bibfield  {journal} {\bibinfo
  {journal} {Phys. Rev. B}\ }\textbf {\bibinfo {volume} {93}},\ \bibinfo
  {pages} {075129} (\bibinfo {year} {2016})}\BibitemShut {NoStop}%
\bibitem [{\citenamefont {McGinley}\ \emph {et~al.}(2017)\citenamefont
  {McGinley}, \citenamefont {Knolle},\ and\ \citenamefont
  {Nunnenkamp}}]{PhysRevB.96.241113}%
  \BibitemOpen
  \bibfield  {author} {\bibinfo {author} {\bibfnamefont {M.}~\bibnamefont
  {McGinley}}, \bibinfo {author} {\bibfnamefont {J.}~\bibnamefont {Knolle}},\
  and\ \bibinfo {author} {\bibfnamefont {A.}~\bibnamefont {Nunnenkamp}},\
  }\bibfield  {title} {\bibinfo {title} {Robustness of majorana edge modes and
  topological order: Exact results for the symmetric interacting kitaev chain
  with disorder},\ }\href {https://doi.org/10.1103/PhysRevB.96.241113}
  {\bibfield  {journal} {\bibinfo  {journal} {Phys. Rev. B}\ }\textbf {\bibinfo
  {volume} {96}},\ \bibinfo {pages} {241113} (\bibinfo {year}
  {2017})}\BibitemShut {NoStop}%
\bibitem [{\citenamefont {Kells}\ \emph {et~al.}(2018)\citenamefont {Kells},
  \citenamefont {Moran},\ and\ \citenamefont {Meidan}}]{PhysRevB.97.085425}%
  \BibitemOpen
  \bibfield  {author} {\bibinfo {author} {\bibfnamefont {G.}~\bibnamefont
  {Kells}}, \bibinfo {author} {\bibfnamefont {N.}~\bibnamefont {Moran}},\ and\
  \bibinfo {author} {\bibfnamefont {D.}~\bibnamefont {Meidan}},\ }\bibfield
  {title} {\bibinfo {title} {Localization enhanced and degraded topological
  order in interacting $p$-wave wires},\ }\href
  {https://doi.org/10.1103/PhysRevB.97.085425} {\bibfield  {journal} {\bibinfo
  {journal} {Phys. Rev. B}\ }\textbf {\bibinfo {volume} {97}},\ \bibinfo
  {pages} {085425} (\bibinfo {year} {2018})}\BibitemShut {NoStop}%
\bibitem [{\citenamefont {Levy}\ and\ \citenamefont
  {Goldstein}(2019)}]{universe5010033}%
  \BibitemOpen
  \bibfield  {author} {\bibinfo {author} {\bibfnamefont {L.}~\bibnamefont
  {Levy}}\ and\ \bibinfo {author} {\bibfnamefont {M.}~\bibnamefont
  {Goldstein}},\ }\bibfield  {title} {\bibinfo {title} {Entanglement and
  disordered-enhanced topological phase in the kitaev chain},\ }\bibfield
  {journal} {\bibinfo  {journal} {Universe}\ }\textbf {\bibinfo {volume} {5}},\
  \href {https://doi.org/10.3390/universe5010033} {10.3390/universe5010033}
  (\bibinfo {year} {2019})\BibitemShut {NoStop}%
\bibitem [{\citenamefont {Huang}\ and\ \citenamefont
  {Yao}(2022)}]{PhysRevB.105.245144}%
  \BibitemOpen
  \bibfield  {author} {\bibinfo {author} {\bibfnamefont {W.}~\bibnamefont
  {Huang}}\ and\ \bibinfo {author} {\bibfnamefont {Y.}~\bibnamefont {Yao}},\
  }\bibfield  {title} {\bibinfo {title} {Intertwined string orders of
  topologically trivial and nontrivial phases in an interacting kitaev chain
  with spatially varying potentials},\ }\href
  {https://doi.org/10.1103/PhysRevB.105.245144} {\bibfield  {journal} {\bibinfo
   {journal} {Phys. Rev. B}\ }\textbf {\bibinfo {volume} {105}},\ \bibinfo
  {pages} {245144} (\bibinfo {year} {2022})}\BibitemShut {NoStop}%
\bibitem [{\citenamefont {Schreiber}\ \emph {et~al.}(2015)\citenamefont
  {Schreiber}, \citenamefont {Hodgman}, \citenamefont {Bordia}, \citenamefont
  {Lüschen}, \citenamefont {Fischer}, \citenamefont {Vosk}, \citenamefont
  {Altman}, \citenamefont {Schneider},\ and\ \citenamefont
  {Bloch}}]{doi:10.1126/science.aaa7432}%
  \BibitemOpen
  \bibfield  {author} {\bibinfo {author} {\bibfnamefont {M.}~\bibnamefont
  {Schreiber}}, \bibinfo {author} {\bibfnamefont {S.~S.}\ \bibnamefont
  {Hodgman}}, \bibinfo {author} {\bibfnamefont {P.}~\bibnamefont {Bordia}},
  \bibinfo {author} {\bibfnamefont {H.~P.}\ \bibnamefont {Lüschen}}, \bibinfo
  {author} {\bibfnamefont {M.~H.}\ \bibnamefont {Fischer}}, \bibinfo {author}
  {\bibfnamefont {R.}~\bibnamefont {Vosk}}, \bibinfo {author} {\bibfnamefont
  {E.}~\bibnamefont {Altman}}, \bibinfo {author} {\bibfnamefont
  {U.}~\bibnamefont {Schneider}},\ and\ \bibinfo {author} {\bibfnamefont
  {I.}~\bibnamefont {Bloch}},\ }\bibfield  {title} {\bibinfo {title}
  {Observation of many-body localization of interacting fermions in a
  quasirandom optical lattice},\ }\href
  {https://doi.org/10.1126/science.aaa7432} {\bibfield  {journal} {\bibinfo
  {journal} {Science}\ }\textbf {\bibinfo {volume} {349}},\ \bibinfo {pages}
  {842} (\bibinfo {year} {2015})},\ \Eprint
  {https://arxiv.org/abs/https://www.science.org/doi/pdf/10.1126/science.aaa7432}
  {https://www.science.org/doi/pdf/10.1126/science.aaa7432} \BibitemShut
  {NoStop}%
\bibitem [{\citenamefont {Lang}\ and\ \citenamefont
  {Chen}(2012)}]{PhysRevB.86.205135}%
  \BibitemOpen
  \bibfield  {author} {\bibinfo {author} {\bibfnamefont {L.-J.}\ \bibnamefont
  {Lang}}\ and\ \bibinfo {author} {\bibfnamefont {S.}~\bibnamefont {Chen}},\
  }\bibfield  {title} {\bibinfo {title} {Majorana fermions in density-modulated
  $p$-wave superconducting wires},\ }\href
  {https://doi.org/10.1103/PhysRevB.86.205135} {\bibfield  {journal} {\bibinfo
  {journal} {Phys. Rev. B}\ }\textbf {\bibinfo {volume} {86}},\ \bibinfo
  {pages} {205135} (\bibinfo {year} {2012})}\BibitemShut {NoStop}%
\bibitem [{\citenamefont {Cai}\ \emph {et~al.}(2013)\citenamefont {Cai},
  \citenamefont {Lang}, \citenamefont {Chen},\ and\ \citenamefont
  {Wang}}]{PhysRevLett.110.176403}%
  \BibitemOpen
  \bibfield  {author} {\bibinfo {author} {\bibfnamefont {X.}~\bibnamefont
  {Cai}}, \bibinfo {author} {\bibfnamefont {L.-J.}\ \bibnamefont {Lang}},
  \bibinfo {author} {\bibfnamefont {S.}~\bibnamefont {Chen}},\ and\ \bibinfo
  {author} {\bibfnamefont {Y.}~\bibnamefont {Wang}},\ }\bibfield  {title}
  {\bibinfo {title} {Topological superconductor to anderson localization
  transition in one-dimensional incommensurate lattices},\ }\href
  {https://doi.org/10.1103/PhysRevLett.110.176403} {\bibfield  {journal}
  {\bibinfo  {journal} {Phys. Rev. Lett.}\ }\textbf {\bibinfo {volume} {110}},\
  \bibinfo {pages} {176403} (\bibinfo {year} {2013})}\BibitemShut {NoStop}%
\bibitem [{\citenamefont {Tezuka}\ and\ \citenamefont
  {Kawakami}(2012)}]{Tezuka2012}%
  \BibitemOpen
  \bibfield  {author} {\bibinfo {author} {\bibfnamefont {M.}~\bibnamefont
  {Tezuka}}\ and\ \bibinfo {author} {\bibfnamefont {N.}~\bibnamefont
  {Kawakami}},\ }\bibfield  {title} {\bibinfo {title} {Reentrant topological
  transitions in a quantum wire/superconductor system with quasiperiodic
  lattice modulation},\ }\href {https://doi.org/10.1103/PhysRevB.85.140508}
  {\bibfield  {journal} {\bibinfo  {journal} {Phys. Rev. B}\ }\textbf {\bibinfo
  {volume} {85}},\ \bibinfo {pages} {140508} (\bibinfo {year}
  {2012})}\BibitemShut {NoStop}%
\bibitem [{\citenamefont {Jordan}\ and\ \citenamefont
  {Wigner}(1928)}]{Jordan1928}%
  \BibitemOpen
  \bibfield  {author} {\bibinfo {author} {\bibfnamefont {P.}~\bibnamefont
  {Jordan}}\ and\ \bibinfo {author} {\bibfnamefont {E.}~\bibnamefont
  {Wigner}},\ }\bibfield  {title} {\bibinfo {title} {Über das paulische
  Äquivalenzverbot},\ }\bibfield  {journal} {\bibinfo  {journal} {Zeitschrift
  für Physik}\ }\textbf {\bibinfo {volume} {47}},\ \href
  {https://doi.org/10.1007/BF01331938} {10.1007/BF01331938} (\bibinfo {year}
  {1928})\BibitemShut {NoStop}%
\bibitem [{\citenamefont {Fendley}(2012)}]{Fendley2012}%
  \BibitemOpen
  \bibfield  {author} {\bibinfo {author} {\bibfnamefont {P.}~\bibnamefont
  {Fendley}},\ }\bibfield  {title} {\bibinfo {title} {Parafermionic edge zero
  modes {inZn}-invariant spin chains},\ }\href
  {https://doi.org/10.1088/1742-5468/2012/11/p11020} {\bibfield  {journal}
  {\bibinfo  {journal} {J. Stat. Mech.}\ }\textbf {\bibinfo {volume} {2012}},\
  \bibinfo {pages} {P11020} (\bibinfo {year} {2012})}\BibitemShut {NoStop}%
\bibitem [{\citenamefont {Calabrese}\ \emph {et~al.}(2012)\citenamefont
  {Calabrese}, \citenamefont {Essler},\ and\ \citenamefont
  {Fagotti}}]{Calabrese_2012}%
  \BibitemOpen
  \bibfield  {author} {\bibinfo {author} {\bibfnamefont {P.}~\bibnamefont
  {Calabrese}}, \bibinfo {author} {\bibfnamefont {F.~H.~L.}\ \bibnamefont
  {Essler}},\ and\ \bibinfo {author} {\bibfnamefont {M.}~\bibnamefont
  {Fagotti}},\ }\bibfield  {title} {\bibinfo {title} {Quantum quench in the
  transverse field ising chain: I. time evolution of order parameter
  correlators},\ }\href {https://doi.org/10.1088/1742-5468/2012/07/P07016}
  {\bibfield  {journal} {\bibinfo  {journal} {Journal of Statistical Mechanics:
  Theory and Experiment}\ }\textbf {\bibinfo {volume} {2012}},\ \bibinfo
  {pages} {P07016} (\bibinfo {year} {2012})}\BibitemShut {NoStop}%
\bibitem [{\citenamefont {Sachdev}(2011)}]{SachdevBook}%
  \BibitemOpen
  \bibfield  {author} {\bibinfo {author} {\bibfnamefont {S.}~\bibnamefont
  {Sachdev}},\ }\href@noop {} {\emph {\bibinfo {title} {Quantum Phase
  Transitions}}}\ (\bibinfo  {publisher} {Cambridge University Press},\
  \bibinfo {year} {2011})\BibitemShut {NoStop}%
\bibitem [{\citenamefont {Lieb}\ \emph {et~al.}(1961)\citenamefont {Lieb},
  \citenamefont {Schultz},\ and\ \citenamefont {Mattis}}]{LSM1961}%
  \BibitemOpen
  \bibfield  {author} {\bibinfo {author} {\bibfnamefont {E.}~\bibnamefont
  {Lieb}}, \bibinfo {author} {\bibfnamefont {T.}~\bibnamefont {Schultz}},\ and\
  \bibinfo {author} {\bibfnamefont {D.}~\bibnamefont {Mattis}},\ }\bibfield
  {title} {\bibinfo {title} {Two soluble models of an antiferromagnetic
  chain},\ }\href {https://doi.org/10.1016/0003-4916(61)90115-4} {\bibfield
  {journal} {\bibinfo  {journal} {Annals of Physics}\ }\textbf {\bibinfo
  {volume} {16}},\ \bibinfo {pages} {407–466} (\bibinfo {year}
  {1961})}\BibitemShut {NoStop}%
\bibitem [{\citenamefont {Selke}(1988)}]{Selke1988}%
  \BibitemOpen
  \bibfield  {author} {\bibinfo {author} {\bibfnamefont {W.}~\bibnamefont
  {Selke}},\ }\bibfield  {title} {\bibinfo {title} {The annni model --
  theoretical analysis and experimental application},\ }\href
  {https://doi.org/https://doi.org/10.1016/0370-1573(88)90140-8} {\bibfield
  {journal} {\bibinfo  {journal} {Physics Reports}\ }\textbf {\bibinfo {volume}
  {170}},\ \bibinfo {pages} {213} (\bibinfo {year} {1988})}\BibitemShut
  {NoStop}%
\bibitem [{\citenamefont {Beccaria}\ \emph {et~al.}(2007)\citenamefont
  {Beccaria}, \citenamefont {Campostrini},\ and\ \citenamefont
  {Feo}}]{Beccaria2007}%
  \BibitemOpen
  \bibfield  {author} {\bibinfo {author} {\bibfnamefont {M.}~\bibnamefont
  {Beccaria}}, \bibinfo {author} {\bibfnamefont {M.}~\bibnamefont
  {Campostrini}},\ and\ \bibinfo {author} {\bibfnamefont {A.}~\bibnamefont
  {Feo}},\ }\bibfield  {title} {\bibinfo {title} {Evidence for a floating phase
  of the transverse annni model at high frustration},\ }\href
  {https://doi.org/10.1103/PhysRevB.76.094410} {\bibfield  {journal} {\bibinfo
  {journal} {Phys. Rev. B}\ }\textbf {\bibinfo {volume} {76}},\ \bibinfo
  {pages} {094410} (\bibinfo {year} {2007})}\BibitemShut {NoStop}%
\bibitem [{\citenamefont {Peschel}\ and\ \citenamefont
  {Emery}(1981)}]{Peschel1981}%
  \BibitemOpen
  \bibfield  {author} {\bibinfo {author} {\bibfnamefont {I.}~\bibnamefont
  {Peschel}}\ and\ \bibinfo {author} {\bibfnamefont {V.~J.}\ \bibnamefont
  {Emery}},\ }\bibfield  {title} {\bibinfo {title} {Calculation of spin
  correlations in two-dimensional ising systems from one-dimensional kinetic
  models},\ }\href {https://doi.org/10.1007/BF01297524} {\bibfield  {journal}
  {\bibinfo  {journal} {Zeitschrift f{\"u}r Physik B Condensed Matter}\
  }\textbf {\bibinfo {volume} {43}},\ \bibinfo {pages} {241} (\bibinfo {year}
  {1981})}\BibitemShut {NoStop}%
\bibitem [{\citenamefont {Allen}\ \emph {et~al.}(2001)\citenamefont {Allen},
  \citenamefont {Azaria},\ and\ \citenamefont {Lecheminant}}]{Allen2001}%
  \BibitemOpen
  \bibfield  {author} {\bibinfo {author} {\bibfnamefont {D.}~\bibnamefont
  {Allen}}, \bibinfo {author} {\bibfnamefont {P.}~\bibnamefont {Azaria}},\ and\
  \bibinfo {author} {\bibfnamefont {P.}~\bibnamefont {Lecheminant}},\
  }\bibfield  {title} {\bibinfo {title} {A two-leg quantum ising ladder: a
  bosonization study of the {ANNNI} model},\ }\href
  {https://doi.org/10.1088/0305-4470/34/21/101} {\bibfield  {journal} {\bibinfo
   {journal} {J. Phys. A: Math. Gen.}\ }\textbf {\bibinfo {volume} {34}},\
  \bibinfo {pages} {L305} (\bibinfo {year} {2001})}\BibitemShut {NoStop}%
\bibitem [{\citenamefont {Beccaria}\ \emph {et~al.}(2006)\citenamefont
  {Beccaria}, \citenamefont {Campostrini},\ and\ \citenamefont
  {Feo}}]{Beccaria2006}%
  \BibitemOpen
  \bibfield  {author} {\bibinfo {author} {\bibfnamefont {M.}~\bibnamefont
  {Beccaria}}, \bibinfo {author} {\bibfnamefont {M.}~\bibnamefont
  {Campostrini}},\ and\ \bibinfo {author} {\bibfnamefont {A.}~\bibnamefont
  {Feo}},\ }\bibfield  {title} {\bibinfo {title} {Density-matrix
  renormalization-group study of the disorder line in the quantum axial
  next-nearest-neighbor ising model},\ }\href
  {https://doi.org/10.1103/PhysRevB.73.052402} {\bibfield  {journal} {\bibinfo
  {journal} {Phys. Rev. B}\ }\textbf {\bibinfo {volume} {73}},\ \bibinfo
  {pages} {052402} (\bibinfo {year} {2006})}\BibitemShut {NoStop}%
\bibitem [{\citenamefont {Sela}\ and\ \citenamefont
  {Pereira}(2011)}]{Sela2011}%
  \BibitemOpen
  \bibfield  {author} {\bibinfo {author} {\bibfnamefont {E.}~\bibnamefont
  {Sela}}\ and\ \bibinfo {author} {\bibfnamefont {R.~G.}\ \bibnamefont
  {Pereira}},\ }\bibfield  {title} {\bibinfo {title} {Orbital multicriticality
  in spin-gapped quasi-one-dimensional antiferromagnets},\ }\href
  {https://doi.org/10.1103/PhysRevB.84.014407} {\bibfield  {journal} {\bibinfo
  {journal} {Phys. Rev. B}\ }\textbf {\bibinfo {volume} {84}},\ \bibinfo
  {pages} {014407} (\bibinfo {year} {2011})}\BibitemShut {NoStop}%
\bibitem [{\citenamefont {White}(1992)}]{White1992}%
  \BibitemOpen
  \bibfield  {author} {\bibinfo {author} {\bibfnamefont {S.~R.}\ \bibnamefont
  {White}},\ }\bibfield  {title} {\bibinfo {title} {Density matrix formulation
  for quantum renormalization groups},\ }\href
  {https://doi.org/10.1103/PhysRevLett.69.2863} {\bibfield  {journal} {\bibinfo
   {journal} {Phys. Rev. Lett.}\ }\textbf {\bibinfo {volume} {69}},\ \bibinfo
  {pages} {2863} (\bibinfo {year} {1992})}\BibitemShut {NoStop}%
\bibitem [{\citenamefont {Schollw\"ock}(2005)}]{Schollwoeck2005}%
  \BibitemOpen
  \bibfield  {author} {\bibinfo {author} {\bibfnamefont {U.}~\bibnamefont
  {Schollw\"ock}},\ }\bibfield  {title} {\bibinfo {title} {The density-matrix
  renormalization group},\ }\href {https://doi.org/10.1103/RevModPhys.77.259}
  {\bibfield  {journal} {\bibinfo  {journal} {Rev. Mod. Phys.}\ }\textbf
  {\bibinfo {volume} {77}},\ \bibinfo {pages} {259} (\bibinfo {year}
  {2005})}\BibitemShut {NoStop}%
\bibitem [{\citenamefont {Schollwöck}(2011)}]{Schollwoeck2011}%
  \BibitemOpen
  \bibfield  {author} {\bibinfo {author} {\bibfnamefont {U.}~\bibnamefont
  {Schollwöck}},\ }\bibfield  {title} {\bibinfo {title} {The density-matrix
  renormalization group in the age of matrix product states},\ }\href
  {https://doi.org/https://doi.org/10.1016/j.aop.2010.09.012.} {\bibfield
  {journal} {\bibinfo  {journal} {Annals of Physics}\ }\textbf {\bibinfo
  {volume} {326}},\ \bibinfo {pages} {96} (\bibinfo {year} {2011})}\BibitemShut
  {NoStop}%
\bibitem [{\citenamefont {Turner}\ \emph {et~al.}(2011)\citenamefont {Turner},
  \citenamefont {Pollmann},\ and\ \citenamefont {Berg}}]{PhysRevB.83.075102}%
  \BibitemOpen
  \bibfield  {author} {\bibinfo {author} {\bibfnamefont {A.~M.}\ \bibnamefont
  {Turner}}, \bibinfo {author} {\bibfnamefont {F.}~\bibnamefont {Pollmann}},\
  and\ \bibinfo {author} {\bibfnamefont {E.}~\bibnamefont {Berg}},\ }\bibfield
  {title} {\bibinfo {title} {Topological phases of one-dimensional fermions: An
  entanglement point of view},\ }\href
  {https://doi.org/10.1103/PhysRevB.83.075102} {\bibfield  {journal} {\bibinfo
  {journal} {Phys. Rev. B}\ }\textbf {\bibinfo {volume} {83}},\ \bibinfo
  {pages} {075102} (\bibinfo {year} {2011})}\BibitemShut {NoStop}%
\bibitem [{\citenamefont {Horodecki}\ \emph {et~al.}(2009)\citenamefont
  {Horodecki}, \citenamefont {Horodecki}, \citenamefont {Horodecki},\ and\
  \citenamefont {Horodecki}}]{RevModPhys.81.865}%
  \BibitemOpen
  \bibfield  {author} {\bibinfo {author} {\bibfnamefont {R.}~\bibnamefont
  {Horodecki}}, \bibinfo {author} {\bibfnamefont {P.}~\bibnamefont
  {Horodecki}}, \bibinfo {author} {\bibfnamefont {M.}~\bibnamefont
  {Horodecki}},\ and\ \bibinfo {author} {\bibfnamefont {K.}~\bibnamefont
  {Horodecki}},\ }\bibfield  {title} {\bibinfo {title} {Quantum entanglement},\
  }\href {https://doi.org/10.1103/RevModPhys.81.865} {\bibfield  {journal}
  {\bibinfo  {journal} {Rev. Mod. Phys.}\ }\textbf {\bibinfo {volume} {81}},\
  \bibinfo {pages} {865} (\bibinfo {year} {2009})}\BibitemShut {NoStop}%
\bibitem [{\citenamefont {personal communication}()}]{dirkcomment}%
  \BibitemOpen
  \bibfield  {author} {\bibinfo {author} {\bibnamefont {personal
  communication}},\ }\href@noop {} {}\BibitemShut {NoStop}%
\bibitem [{\citenamefont {Khemani}\ \emph {et~al.}(2016)\citenamefont
  {Khemani}, \citenamefont {Pollmann},\ and\ \citenamefont
  {Sondhi}}]{Khemani2016}%
  \BibitemOpen
  \bibfield  {author} {\bibinfo {author} {\bibfnamefont {V.}~\bibnamefont
  {Khemani}}, \bibinfo {author} {\bibfnamefont {F.}~\bibnamefont {Pollmann}},\
  and\ \bibinfo {author} {\bibfnamefont {S.~L.}\ \bibnamefont {Sondhi}},\
  }\bibfield  {title} {\bibinfo {title} {Obtaining highly excited eigenstates
  of many-body localized hamiltonians by the density matrix renormalization
  group approach},\ }\href {https://doi.org/10.1103/PhysRevLett.116.247204}
  {\bibfield  {journal} {\bibinfo  {journal} {Phys. Rev. Lett.}\ }\textbf
  {\bibinfo {volume} {116}},\ \bibinfo {pages} {247204} (\bibinfo {year}
  {2016})}\BibitemShut {NoStop}%
\end{thebibliography}%

\end{document}